\documentclass[twocolumn,floatfix,pra,aps,showpacs,amsmath,superscriptaddress]{revtex4-1}
\usepackage{epsfig,graphicx,tabularx,epstopdf,physics,color,bm,amssymb}

\usepackage[utf8]{inputenc}
\begin{document}
\title{Blocked populations in ring-shaped optical lattices}
\author{M. Nigro}
\author{P. Capuzzi} 
\author{D. M. Jezek} 
\affiliation{Universidad de
  Buenos Aires, Facultad de Ciencias Exactas y Naturales, Departamento
  de F\'{i}sica, Buenos Aires, Argentina} 
\affiliation{IFIBA,
  CONICET-UBA, Pabell\'on 1, Ciudad Universitaria, 1428 Buenos Aires,
  Argentina } 
\date{\today}
\begin{abstract}

  We study a special dynamical regime of a Bose-Einstein condensate in
  a ring-shaped lattice where the populations in each site remain
  constant during the time evolution.  The states in this regime are
  characterized by equal occupation numbers in alternate wells and
  non-trivial phases, while the phase differences between neighboring
  sites evolve in time yielding persistent currents that oscillate
  around the lattice.  We show that the velocity circulation around
  the ring lattice alternates between two values determined by the
  number of wells and with a specific time period that is only driven
  by the onsite interaction energy parameter. In contrast to the
  self-trapping regime present in optical lattices, the occupation
  number at each site does not show any oscillation and the particle
  imbalance does not possess a lower bound for the phenomenon to
  occur.  These findings are predicted with a multimode model and
  confirmed by full three-dimensional Gross-Pitaevskii simulations
  using an effective onsite interaction energy parameter.

\end{abstract}
\pacs{03.75.Lm, 03.75.Hh, 03.75.Kk}

\maketitle
\section{Introduction}

The self-trapping phenomenon has been extensively studied in
double-well systems by means of a two-mode model
\cite{smerzi97,ragh99,anan06,jia08,mele11,abad11,mauro17,doublewell},
and experimentally observed by Albiez {\it et al.}  \cite{albiez05,
  thesisAlbiez,thesisGati}. In this regime the population in one site
remains higher than the one in the other well over all the evolution.
This imbalance of particles performs oscillations around the non
vanishing mean value, whereas the phase difference between the sites
exhibits a running phase behavior.  Theoretical studies of this
phenomenon has been also carried out by several authors in extended
regular lattices \cite{Anker2005,Wang2006,optlat,stlastoplat}. More
recently, the study of self-trapping has also been addressed in
ring-shaped optical lattices \cite{cuatropozos06,arwas2014,mauro4p}.
Such works treat three- and four-well systems. In Refs.\
\cite{cuatropozos06,arwas2014} the dynamics has been investigated
through a multimode (M) model that utilized {\em ad-hoc} values for
the hopping and onsite energy parameters.  Whereas in
Ref. \cite{mauro4p}, such parameters have been extracted from a
mean-field approach using three-dimensional localized ``Wannier-like''
(WL) onsite functions and including an effective onsite interaction
energy parameter \cite{jezek13a,jezek13b}.  For large filling numbers,
the inclusion of such a realistic interaction parameter has been shown
to be crucial for the accurate description of the dynamics, yielding a
sizable change on the time periods respect to those obtained by the
standard model. In contrast, for filling number around unity
mean-field approaches are not applicable and hence other microscopic
methods have to be used \cite{Kolovski2006,Gallemi2016}. In Ref.\
\cite{Burchianti2017}, it has been demonstrated that an effective
interaction can also be extracted from the Bogoliubov excitations in
the case of the Josephson regime.  A systematic study of the
self-trapping regime and the crossover to the Josephson oscillations
in four-well systems including non-symmetric configurations has been
developed in Ref.\ \cite{mauro4p}. It is worthwhile noticing that the
dynamics in multiple well condensates constitutes a promising area
provided that successful efforts have been performed to experimentally
construct ring-shaped optical lattices \cite{hen09}.

In this work we demonstrate theoretically the existence of a dynamical
regime that exhibits a novel behavior. If the number of wells of the
lattice is a multiple of four, there exists a family of nonstationary
states with constant site populations and special non-trivial
phases. These states could be regarded as a special variation of a ST
regime where, in contrast to that observed in two- and multiple-well
condensates \cite{smerzi97,ragh99,mauro17,mauro4p}, the population
imbalance between neighboring sites can be arbitrarily low and do not
exhibit any oscillation in time. For such states the M model order
parameter can be expressed as a linear combination of particular
degenerate Gross-Pitaevskii (GP) stationary states. However, due to
the nonlinear nature of the GP equation, the states are
nonstationary. The dynamics of these states is governed only by the
onsite interaction energy parameter. We explicitly show that the
angular momentum exhibits a simple oscillating behavior and that the
velocity circulation around the ring alternates periodically between
values $-N_c/4$ and $N_c/4$, being $N_c$ the number of weakly linked
condensates. A goal of this work is to obtain an analytical expression
for such a time period which involves only the imbalance and the
effective interaction parameter.  By comparing the evolution of the
phase differences obtained through GP simulations for a four-well
system and with the M model, we can establish the accuracy of such a
parameter. The existence of these states is confirmed numerically by
means of full three-dimensional Gross-Pitaevskii (GP) simulations
showing a perfect accordance to the M model predictions for several
population imbalances. Furthermore, a Floquet stability analysis
confirms that for the imbalances studied here the dynamics turns out
to be regular.

The paper is organized as follows. In Sec. \ref{multimode} we briefly
review the main concepts of the multimode model. In particular, we
rewrite the equations of motion which include an effective onsite
interaction parameter \cite{mauro4p} and we outline the construction
of the localized states in terms of the GP stationary ones.  In
Sec. \ref{trap4} we describe the specific four-well system used in the
numerical simulations.  Section \ref{states} is devoted to study the
properties of these states with blocked occupation numbers.  As a
first step we introduce a continuous family of states corresponding to
fixed points of the M model in the phase diagram defined by the
populations and phase differences. On the other hand, we
demonstrate that they turn out to be quasi-stationary solutions of the
GP equation.  Such states are defined with a particular combination of
phases which give rise the non-stationary blocked-occupation-number
(BON) states.  Secondly, we show that these BON states describe closed
orbits in the phase diagram whose time period is solely determined by
the onsite interaction energy.  By performing GP numerical simulations
with a four-well potential we analyze the hidden dynamics which
includes variations of density in the interwell regions, oscillations
of the velocity field circulation, and an active vortex dynamics.  We
end this section with a study of the Floquet stability of the BON
states and a proposed experimental test. In Sec. \ref{larger} we show
how to generalize the previous results for systems with larger number
of sites. To conclude, a summary of our work is presented in
Sec. \ref{sum} and the definition of the parameters employed in
the equations of motion are gathered in the Appendix.

\section{Multimode model}\label{multimode}

The equations of motion of the multimode model has been previously
studied both for multiple-well systems in general
\cite{cat11,jezek13b} and also in the case of a four-well system
\cite{cuatropozos06,mauro4p}. Here, we only review its main
ingredients, focusing in the definition of their localized states
extracted from the stationary solutions of the GP equations.

\subsection{Multimode model equations of motion including interaction-driven corrections}

Using the multimode model  order parameter,
\begin{equation}
  \psi_M (t,{\mathbf r}) = \sum_k \,  b_k(t)  \,  w_k ({ r, \theta,z})
  \,,
\label{orderparameter}
\end{equation}
written in terms of three-dimensional WL functions localized at the
$k$-site, $w_k(\mathbf{r})$ \cite{mauro4p}, one obtains the equations
of motion for the time dependent coefficients
$b_k(t)= e^{i \phi_k} \, |b_k| $, by replacing the order parameter in
the time dependent GP equation.  The $ 2 N_c $ real equations, written
in terms of the populations $ n_k =|b_{k}|^2 = N_k / N $ and phase
differences $ \varphi_k= \phi_k - \phi_{k-1}$ including effective
onsite interaction corrections \cite{mauro4p}, are

\begin{eqnarray}
 \hbar\,\frac{dn_k}{dt}& = & - 2 J \left[ \sqrt{n_k \, n_{k+1}} \, \sin\varphi_{k+1} 
- \sqrt{n_k \, n_{k-1} } \, \sin\varphi_k \right ]\nonumber\\
&-&  2 F \left[ \sqrt{n_k \, n_{k+1} } (n_k + n_{k+1} ) \, \sin\varphi_{k+1} \right. \nonumber\\
&-& \left.\sqrt{n_k \, n_{k-1} } (n_k + n_{k-1} ) \, \sin\varphi_k\right] \, ,
\label{ncmode1hn}
\end{eqnarray}
\begin{eqnarray}
 \hbar\,\frac{d\varphi_k}{dt} & = &   ( n_{k-1} -n_{k}) N  U_{\text{eff}}  \nonumber\\  
&-& \alpha (  n_{k-1} - n_{k}) N U \left[   N_c (n_{k-1}+  n_{k})-2 \right]  \nonumber\\
&-&  J \left[ \left(\sqrt{\frac{n_k}{ n_{k-1}}} 
- \sqrt{\frac{n_{k-1} }{ n_k}}\,\right) \, \cos\varphi_k \right.\nonumber\\
&+& \left. \sqrt{\frac{n_{k-2}}{ n_{k-1} }} \, \cos\varphi_{k-1} 
- \sqrt{\frac{n_{k+1} }{ n_k}} \, \cos\varphi_{k+1} \right]\nonumber\\
&-&  F  \left[ \left( n_k \sqrt{\frac{n_k}{ n_{k-1} }} - n_{k-1} \sqrt{\frac{n_{k-1}}{ n_k}}\,\right)
 \, \cos\varphi_k   \right.\nonumber\\
 &+&  \left( 3\, \sqrt{n_{k-2} \, n_{k-1}} + n_{k-2} \sqrt{\frac{n_{k-2}}{ n_{k-1}}}\,\right) 
 \, \cos\varphi_{k-1} \nonumber\\
&-& \left.\left(  3\, \sqrt{n_{k+1} \, n_k} + n_{k+1} \sqrt{\frac{n_{k+1}} { n_k}}\,\right)  \, 
\cos\varphi_{k+1}\right], \nonumber \\
\label{ncmode2hn}
\end{eqnarray}
where $U_{\text{eff}}= f_{3D} U$. The definitions of the tunneling
parameters $J$ and $F$, and of the onsite interaction energy parameter
$U$ are given in the Appendix.  The coefficient $f_{3D}= 1-\alpha$ is
obtained from the slope of the onsite interaction energy as function
of $\Delta N_k - N/N_c$.  As shown in Refs. \cite{jezek13b,mauro4p},
the introduction of $f_{3D}$ is crucial for obtaining an accurate
dynamics.  From this system of equations only $2N_c- 2$ are
independent since the variables must fulfill $\sum_k n_k=1$ and
$\sum_k \varphi_k=0$.

\subsection{ Localized states}

In previous works, we have described in detail the method for
obtaining the localized states in terms of GP stationary states
\cite{mauro4p,cat11,jezek13b}.  Summarizing, first the stationary
states $\psi_n( r, \theta, z )$ are obtained as the numerical
solutions of the three-dimensional GP equation \cite{gros61} with
different winding numbers $n$, with $n$ restricted to the values
$-[(N_c-1)/2]\leq n \leq [N_c/2]$ \cite{je11} for large barrier
heights \cite{cat11}. Since the $\psi_n$ are orthogonal for different
$n$ \cite{cat11,jezek13b}, one can define orthogonal WL functions
localized on the $k$-site by the following expression:
\begin{equation}
w_k({ r, \theta, z })=\frac{1}{\sqrt{N_c}} \sum_{n} \psi_n({ r, \theta, z})
 \, e^{-i n\theta_k } \,,
\label{wannier}
\end{equation}
with $\theta_k = 2 \pi k / N_c$ for $-[(N_c-1)/2]\leq k\leq [N_c/2]$.
A discussion of how to choose the global phases of
$ \psi_n({ r, \theta, z})$ in order to achieve the maximum
localization of $w_k$ is given in Ref. \cite{mauro4p}.

In its turn, the stationary wavefunctions can be written in terms of
the localized WL wavefunctions in Eq.~(\ref{wannier}) as
\begin{equation}
  \psi_n({r,\theta,z}) = \frac{1}{\sqrt{N_c}}\sum_{k} w_k({r,\theta, z}) e^{i n\, k\,2\pi/N_c}.
\end{equation}
For the four-well problem, $N_c=4$, the states with $n=\pm 1$ are
degenerate and can be regarded as vortex-antivortex states since
$\psi_{\pm 1}=\frac{1}{2}\sum\limits_kw_k e^{\pm i\frac{\pi}{2} k}$
have opposite circulation. It is worthwhile remarking that as the GP
equation is nonlinear, linear combinations of degenerate stationary
states, as , e.g., the vortex-antivortex states, are in general
nonstationary.

\section{ The system, trapping potential and parameters  }\label{trap4}

Although the states investigated in this work also exist for larger
number of wells, in our numerical simulations we will consider a
four-well ring-shaped trapping potential given by
\begin{equation}
V_{\text{trap}}({\bf r} ) =  \frac{ m }{2 }    \left[
\omega_{r}^2  r^2  
+ \omega_{z}^2  z^2 \right] 
+   
V_b \left[  \cos^2(\pi x/q_0)+   
 \cos^2(\pi y/q_0)\right],
\label{eq:trap4}
\end{equation}
where $r^2=x^2+y^2$ and $m$ is the atom mass. The harmonic frequencies
are given by $ \omega_{r}= 2 \pi \times 70 $ Hz and
$ \omega_{z}= 2 \pi \times 90 $ Hz, and the lattice parameter is
$ q_0= 5.1 \mu$m.  Hereafter, time and energy will be given in units
of $\omega_r^{-1}$ and $\hbar\omega_r$, respectively. The length will
be given in units of the radial oscillator length
$l_r=\sqrt{\hbar/(m\omega_r)}\simeq 1.3\,\mu$m. We also fix the
barrier height parameter at $V_b = 25 \hbar\omega_r $ and the number
of particles to $ N=10^4 $.

For a system of Rubidium atoms in the above configuration we have
obtained the following multimode parameters, the hopping
$ J=-6.60\times 10^{-4} \hbar \omega_r $, the interaction driven
hopping parameter $F= 2.08\times 10^{-3} \hbar\omega_r$, the onsite
interaction energy $U= 3.16\times 10^{-3} \hbar\omega_r$, and the
effective onsite interaction energy
$U_{\mathrm{eff}}=2.27 \times10^{-3}\hbar\omega_r$, being
$\alpha= 0.28 $. We will numerically solve the GP equation on a grid
of up to $512\times 512\times 256$ points and using a second-order
split-step Fourier method for the dynamics with a time step of
$\Delta t = 10^{-4}\omega_r^{-1}$. For more details see
Ref. \cite{mauro4p}.

\section{ The states }\label{states}

In this section we will first analyze a set of stationary points of
the M model with equally populated sites whose associated order
parameters are in general not exact GP stationary states. These states
shall be called peculiar. In a second step, we shall show that for
states with conveniently chosen initial occupation numbers and the
same distribution of initial phases as the peculiar states, the
populations remain blocked during all the evolution. The properties of
such BON states shall be studied next.

\subsection{ Peculiar stationary states }\label{peculiar}

\begin{figure}
  \begin{center}
    \begin{tabular}{cc}
\includegraphics[width=0.55\columnwidth,clip=true]{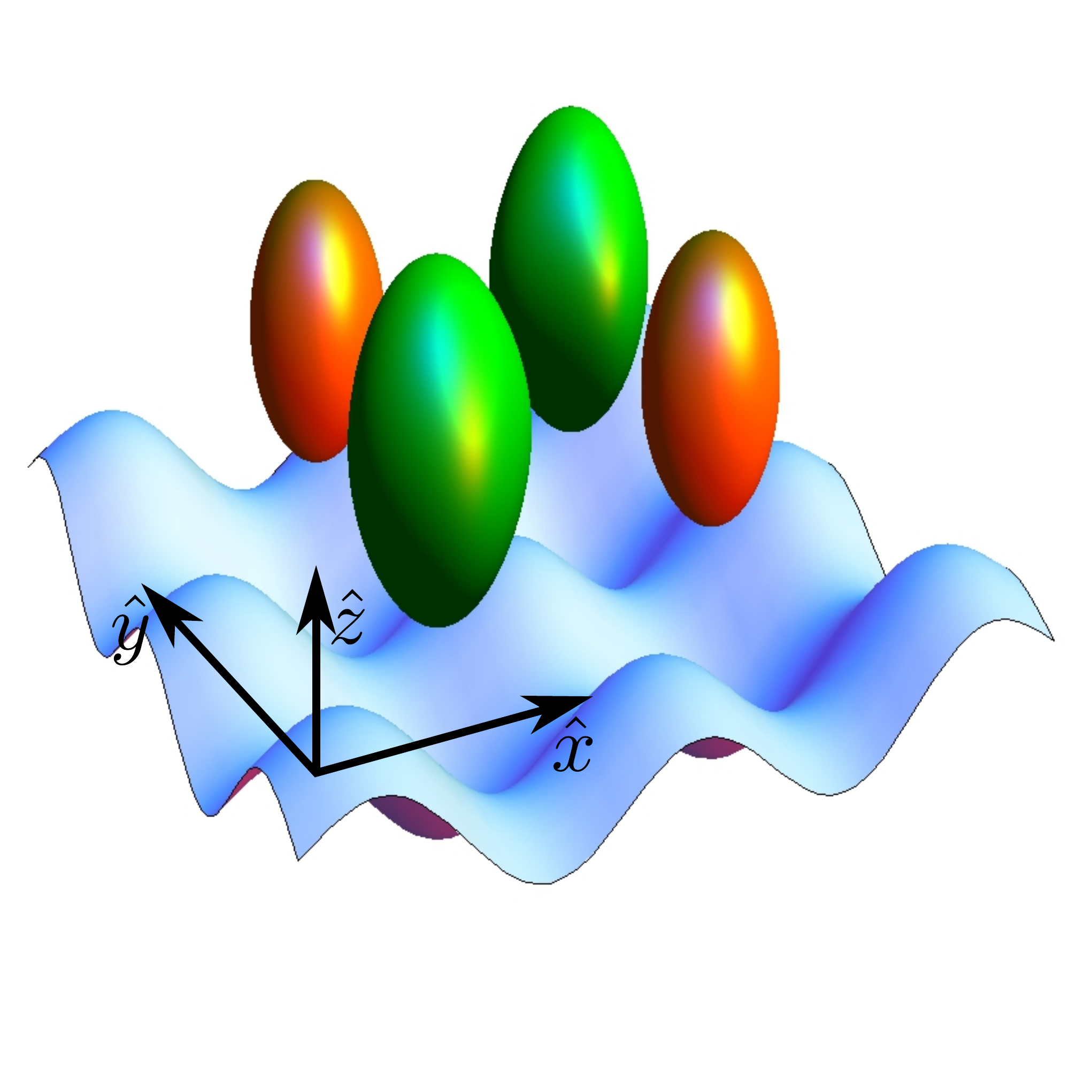} &
\includegraphics[width=0.42\columnwidth,clip=true]{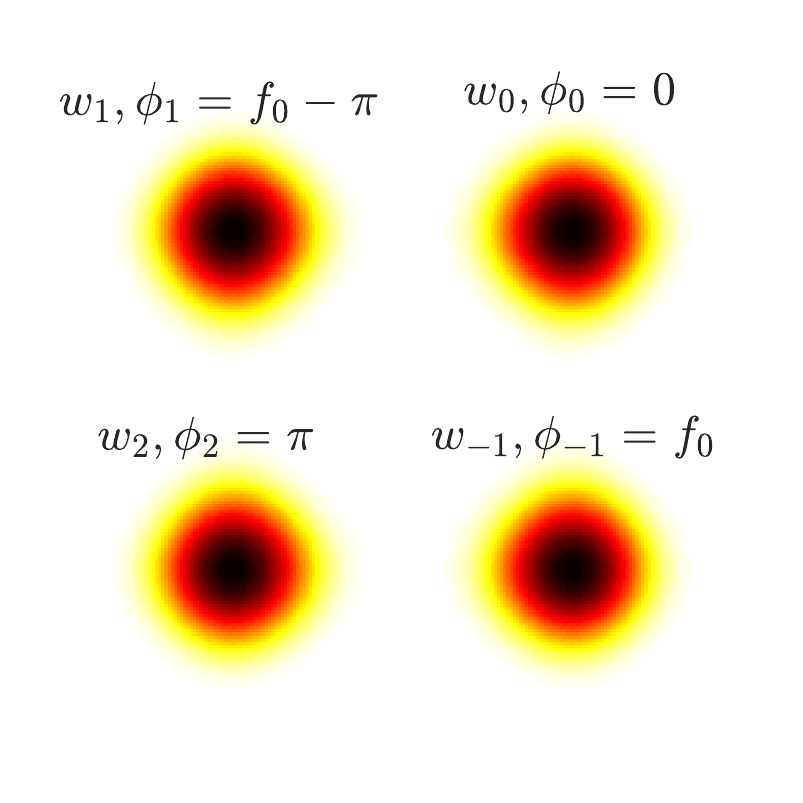}\\
      (a) &(b)
\end{tabular}
\end{center}
\caption{\label{fig:rho3D} (color online) (a): Schematic
  three-dimensional states density and the trapping potential of the
  four-site system (in arbitrary units).  (b): Localized states $ w_k$
  at the plane $z=0$, also the peculiar set of initial values of the
  phases are indicated.}
\end{figure}

The GP stationary states used for constructing the multimode model
give rise to stationary points in the M model.  However, in addition
to these standard points, we found a peculiar set of stationary points
in a condensate with $N_c=4l$ sites. These states are defined by
$ | b_k| = 1/\sqrt{N_c} $ and the following local phases:
$ \phi_0 =0$, $ \phi_{ 1} = f_0 - \pi $, $ \phi_2 =\pi$, and
$ \phi_{ -1} = f_0 $ for a four-well trap ($l=1$). Whereas for larger
$l$, the sequence of phases is repeated $l$ times along the ring.  We
refer to these states as peculiar because $f_0$ could take any value,
so that instead of having isolated points in the phase diagram we have
a continuous family of stationary points parametrized by $f_0$.  This
family contains the two stationary points $f_0= \pm \pi/2$ which
correspond to singly-quantized vortex states, namely, GP stationary
states with winding numbers $\pm 1$. In Fig. \ref{fig:rho3D} (a) a
scheme of the trap and the condensate is depicted qualitatively
showing states with different populations, and in Fig. \ref{fig:rho3D}
(b) the localized WL function in the $z=0$ plane are shown together
with the peculiar initial phases.

We further investigate if the peculiar order parameter could also be
another stationary solution of the GP equation \cite{gros61},
\begin{equation}
 \left[ -\frac{ \hbar^2}{2m} \nabla^2 + V_{\text{trap}} +
g \, |\psi(\mathbf{r})|^2   \right] \psi(\mathbf{r})= \mu  \, \psi(\mathbf{r}), 
\label{gp}
\end{equation}
where $\mu $ is the chemical potential and $g=4\pi \hbar^2 a /m$ is the
interaction strength among atoms with $a$ being their $s$-wave
scattering length.

The normalized-to-unity order parameter associated to the peculiar
points reads,
\begin{equation}
  \psi_M ({\mathbf r},t)  =   \frac{1}{2}  [ w_0(\mathbf{r})-w_2(\mathbf{r})] -   \frac{1}{2}  
 [{w}_1(\mathbf{r})-{w}_{-1}(\mathbf{r})]  e^{i  f_0}         
\label{orderwi0}
\end{equation}
which in terms of GP stationary states can be written as
\begin{equation}
  \psi_M ({\mathbf r})   =   \frac{ 1}{2}   \left[ ( 1    +  i  e^{i  f_0}     )  
   \psi_1({\mathbf r}) 
    + (1    -  i e^{i   f_0}    )     \psi_{-1}({\mathbf r})   \right]  \, .   
\label{gcuatrod}
\end{equation}
The peculiar states are therefore a superposition of vortex states
with opposite circulation.
Since the states $ \psi_{1}({\mathbf r})$ and
$\psi_{-1}( {\mathbf r})$ have the same chemical potential
$\mu_1=\mu_{-1}$ and verify
$ \psi_{1}({\mathbf r}) =\psi^*_{-1}({\mathbf r})$, applying the GP
equation (\ref{gp}) to $\psi_M$ we obtain,
\begin{multline}
 \left[ -\frac{ \hbar^2}{2m} \nabla^2 + V_{\text{trap}} +
g \, |\psi_M (\mathbf{r})|^2   \right] \psi_M ({\mathbf r})  =   \mu_1  \,  \psi_M ({\mathbf r})  \\  
 -  g N \cos(f_0) \Im\left( \psi_{1}^2({\mathbf r})\right) \left[ \Re ( \psi_{1} ) - \Im (\psi_{1}) e^{i f_0} \right] 
\label{gcuatrod2}
\end{multline}
where in addition, we have that
\begin{equation}
\Im\left (\psi_{1}^2({\mathbf r})\right) = \frac{1}{4}
 [w_0({\mathbf r})-w_2({\mathbf r})] \, [ w_1({\mathbf r})-w_{-1}({\mathbf r})]   
\label{gcuatrod3}
\end{equation}
is almost vanishing if the WL functions are well localized as it is in
the present case. Therefore, these peculiar stationary points can be
regarded as quasi-stationary solutions of the GP equation.

For the particular case of $f_0= \pm \pi/2$, the second term of the
right hand side of Eq. (\ref{gcuatrod2}) vanishes and thus the order
parameter is an exact solution of the GP equation. Otherwise, a
general value of $f_0$ generates an entire continuous family of states
that shows a collective motion independent of time.  The
$\psi _{\pm 1}$ stationary solutions can be regarded as particular
cases of Eq. (\ref{gcuatrod}) with maximum angular momentum.  On the
other hand, for $f_0=0$ we have
$\psi_M ({\mathbf r}) = \frac{ 1}{2} \left[ ( 1 + i ) \psi_1({\mathbf
    r}) + (1 - i ) \psi_{-1}({\mathbf r}) \right]$ which is real,
hence its angular momentum is zero.  Nevertheless an active vortex
dynamics is present, due to the nonzero circulation of
$\psi _{\pm 1}$.  The same holds for $f_0=\pi$.

We want to remark that such a family of stationary solutions of the M
model does not necessarily exist in ring lattices with an arbitrary
number of wells as it can be straightforwardly deduced from the
dynamical equations (\ref{ncmode1hn}) and (\ref{ncmode2hn}). For
example, for $N_c=3$ even though the degenerate stationary states with
winding numbers $n= \pm 1$ are also present, the corresponding
stationary points in the phase diagram only exist as isolated points.

\subsection{ Nonstationary BON states}

\begin{figure}
\begin{tabular}{cc}
\includegraphics[width=0.5\columnwidth,clip=true]{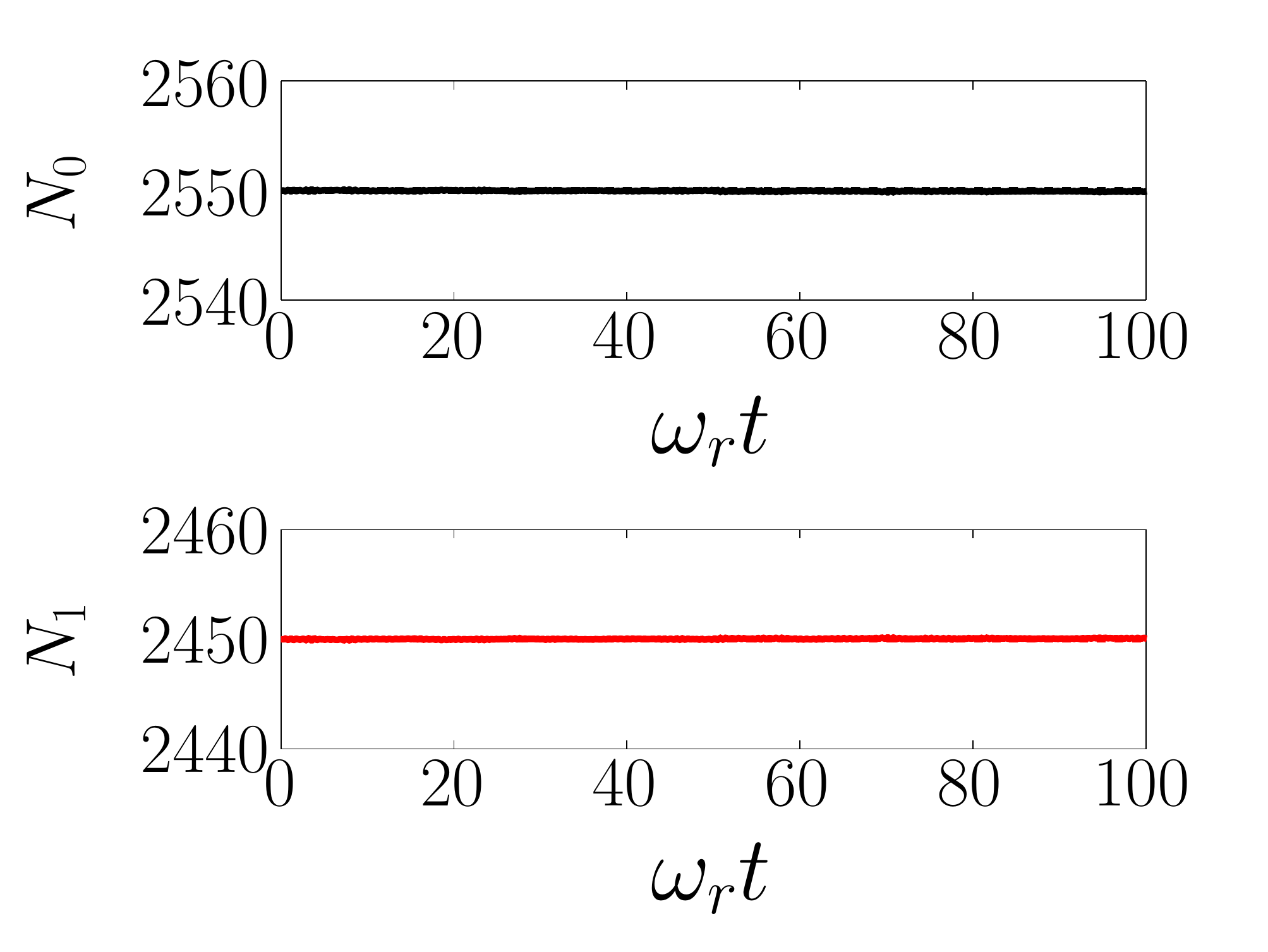}&
\includegraphics[width=0.5\columnwidth,clip=true]{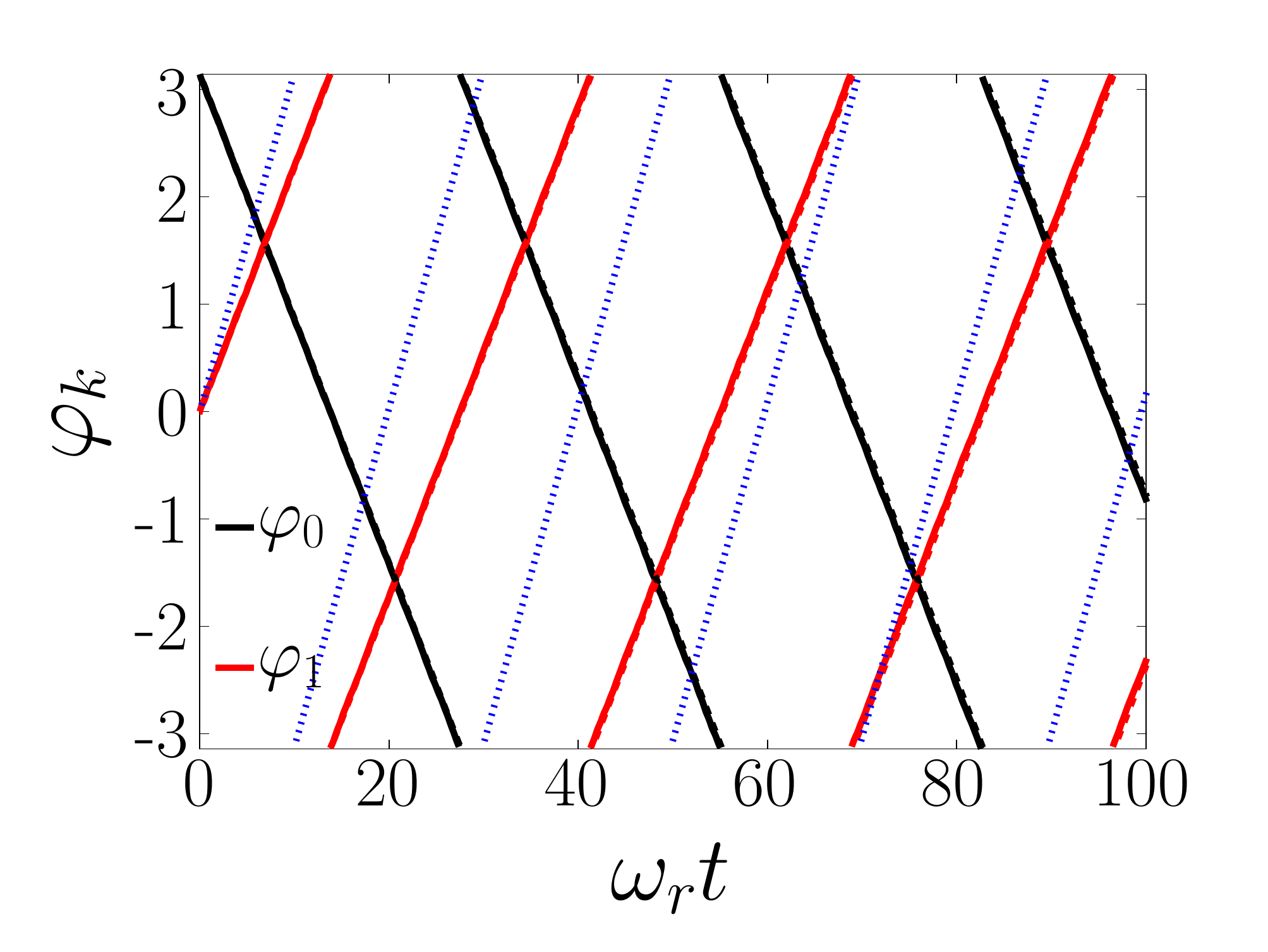} \\
\includegraphics[width=0.5\columnwidth,clip=true]{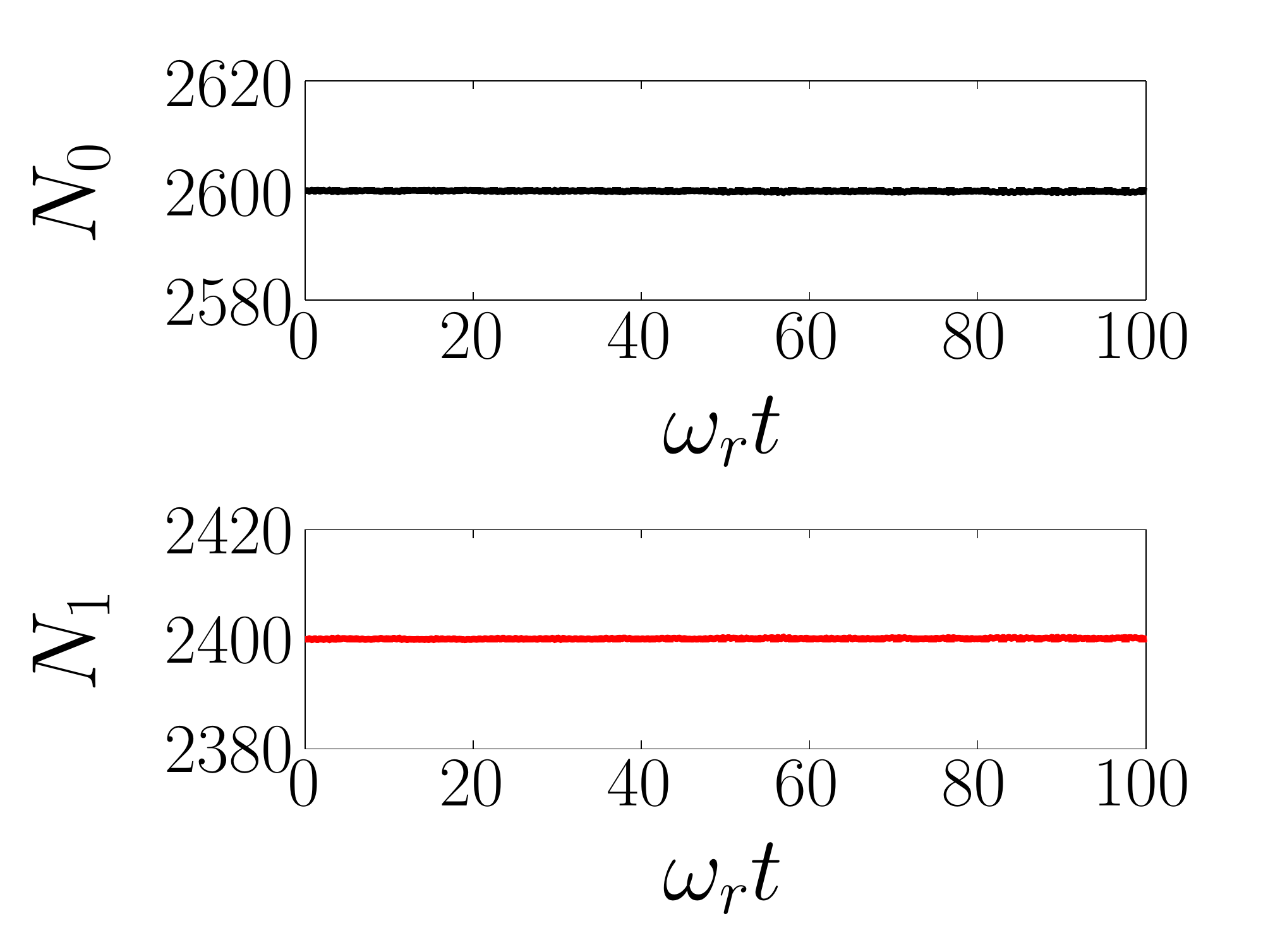}&
\includegraphics[width=0.5\columnwidth,clip=true]{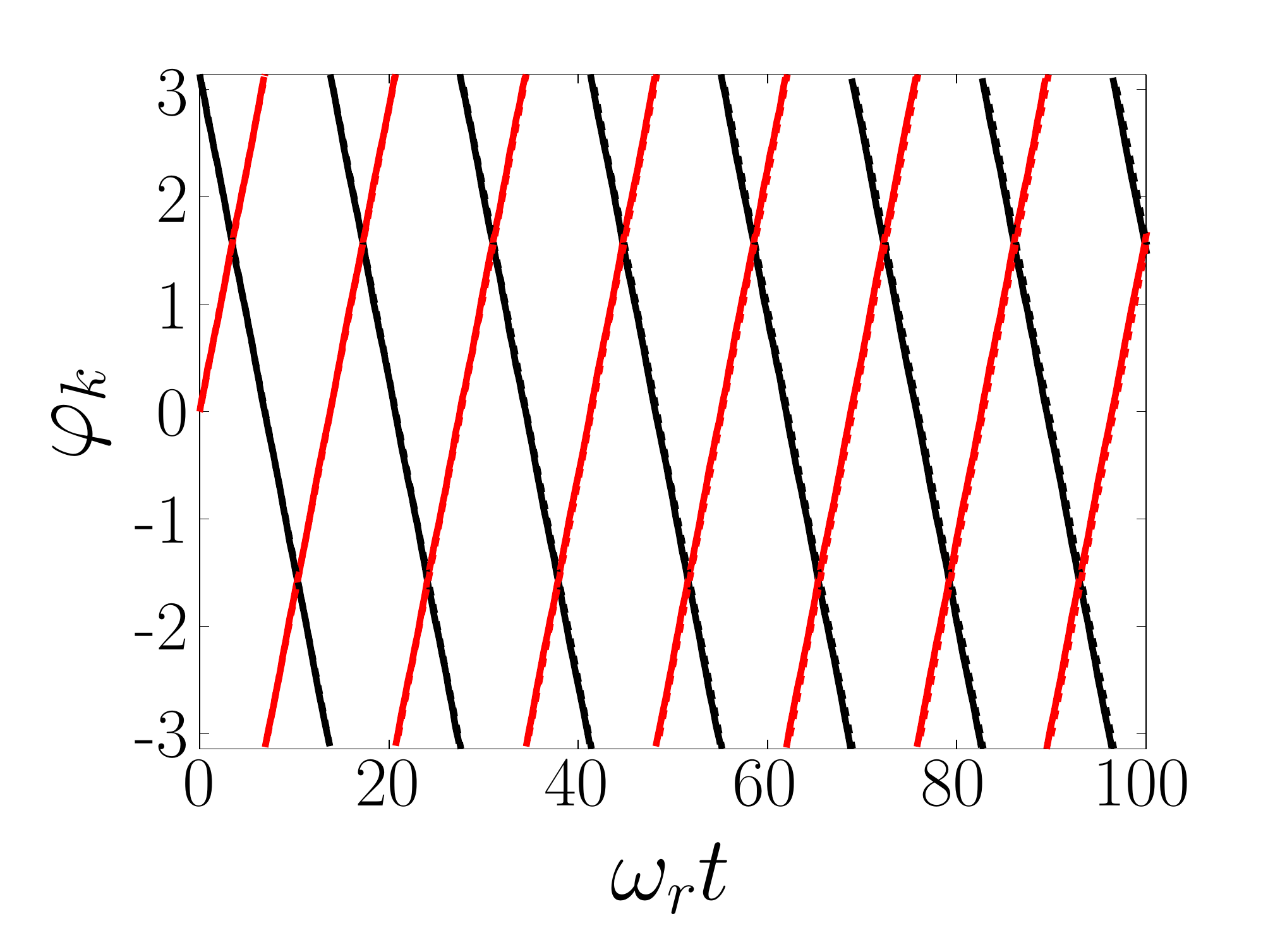} \\
\includegraphics[width=0.5\columnwidth,clip=true]{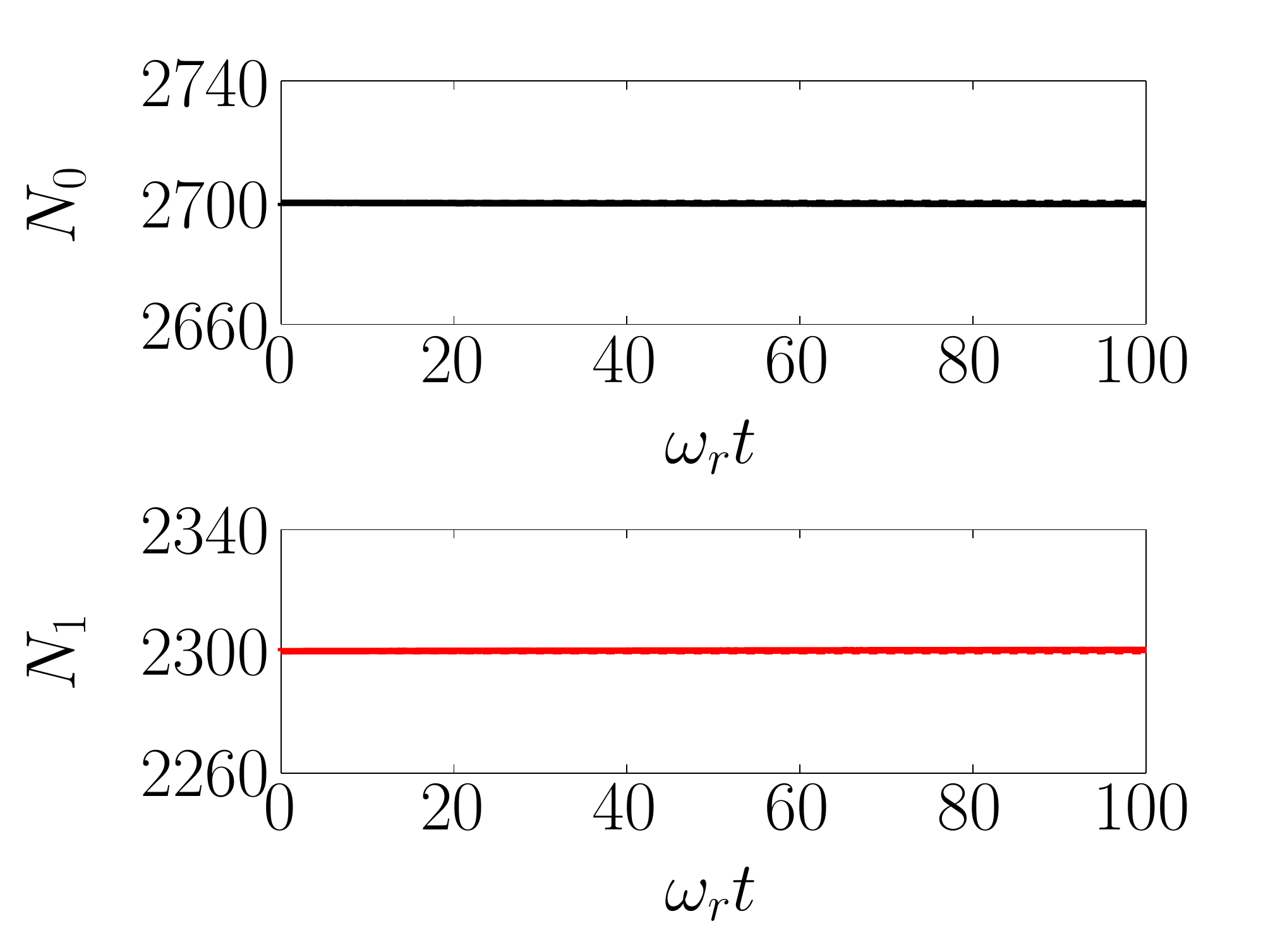}&
\includegraphics[width=0.5\columnwidth,clip=true]{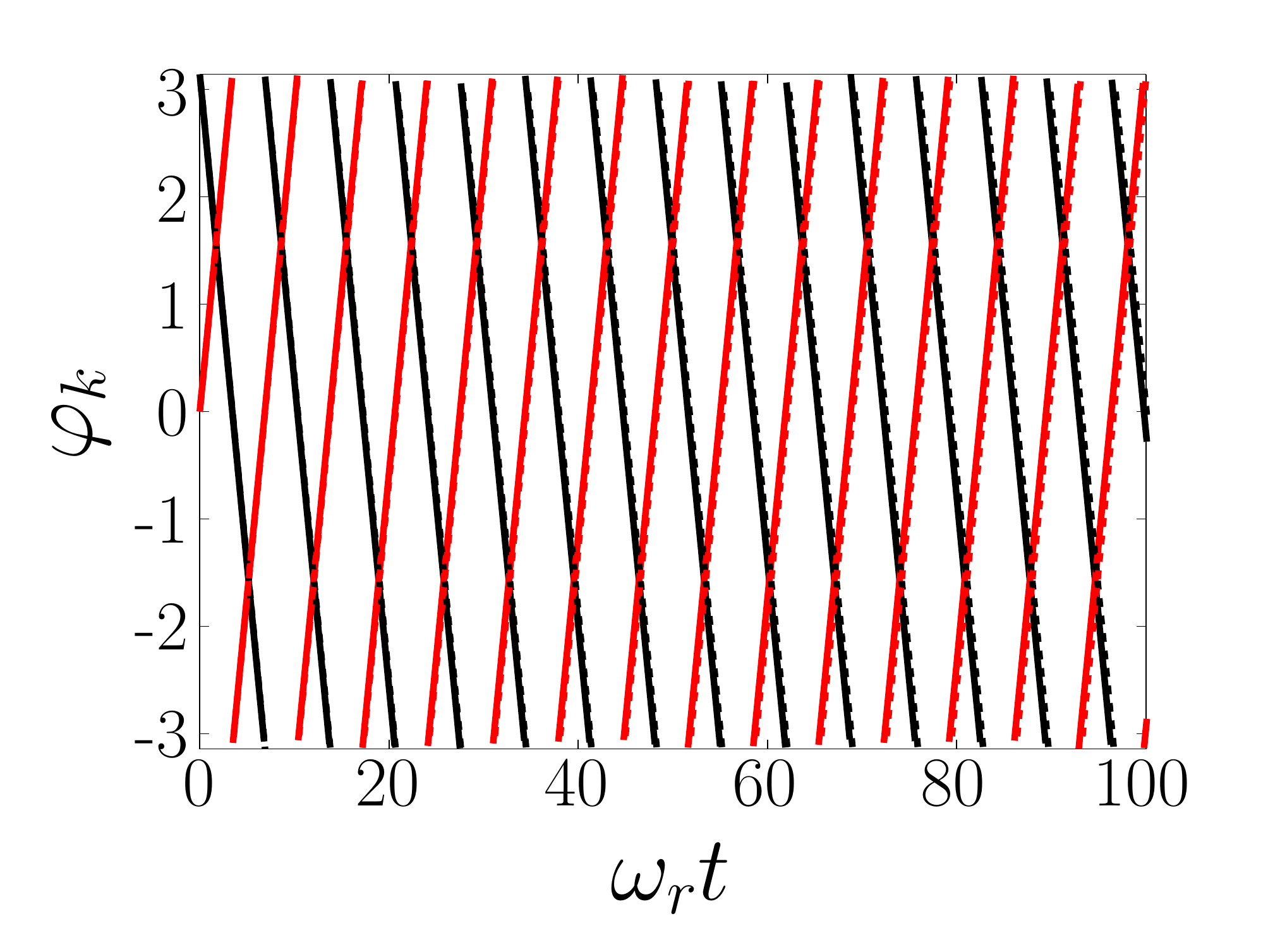}
\end{tabular}
\caption{\label{fig:popu} Populations (left column) and phase
  differences (right column) as function of time, from top to bottom
  for $\Delta N=100$, $200$, and $400$ with $f_0(0)= -\pi$. The solid
  lines correspond to GP simulations and the dashed lines to the
  M model. Black and red lines correspond to $N_k$ and $ \varphi_k$ with
  $k=0$ and $k=1$, respectively. The dotted blue lines in the top
  right panel illustrates the prediction of the M model with the bare
  $U$. }
\end{figure}

When the numbers of particles of alternate sites are equal and the
phases maintain their peculiar relation: $ \phi_0 =0$,
$ \phi_{ 1} = f_0 - \pi $, $ \phi_2 =\pi$, and $ \phi_{ -1} = f_0 $,
the site populations do not evolve. This condition gives rise to very
special dynamical states where $f_0$ becomes time dependent.  This is
shown in Fig. \ref{fig:popu} where we compare the evolution of the
occupation numbers and phase differences using full three-dimensional
GP simulations with the dynamics arising from the M model.  The
selected initial population differences, from top to bottom, are
$\Delta N =100$, $200$, and $400$, where $\Delta N= N_0 - N_1 $ and
$f_0(0)= - \pi $.  From Fig. \ref{fig:popu} it may be seen that in all
cases the number of particles in each site remains fixed in time,
whereas their phase differences evolve faster for larger
imbalances. We have further investigated the GPE dynamics for
imbalances up to $\Delta N = 3\times 10^3$ and verified that the
populations remain constant within $0.1\%$ accuracy.

These family of states bear some resemblance to self-trapped states;
however, there are many important differences compared to the
well-known ST dynamics in double well potentials. First of all, the
population imbalance can be arbitrarily small. Instead, to reach these
states, it is only necessary to achieve the peculiar phases described
above, being $f_{0}(0)$ an arbitrary value (given that
$n_{k}=n_{k+2}$). Second, the hopping parameters $J$ and $F$ play no
role in the dynamics, hence we cannot associate the emergence of the
BON states to the small enough tunneling energy splitting like in the
ST regime in two wells \cite{albiez05}.
 
The BON state normalized to unity  reads,
\begin{equation}
  \psi_M ({\mathbf r},t)  =   \sqrt{n_0} [
  {w}_0(\mathbf{r})-{w}_2(\mathbf{r})]  -     \sqrt{n_1} 
  [{w}_1(\mathbf{r})-{w}_{-1}(\mathbf{r})]  e^{i  f_0(t)}         
\label{bon1}
\end{equation}
which written in terms of GP stationary states yields,
\begin{multline}
  \psi_M ({\mathbf r})   =     \left(   \sqrt{n_0}   +  i  \sqrt{n_1}  e^{i  f_0(t)}     \right)  
   \psi_1({\mathbf r}) 
    \\ + \left( \sqrt{n_0}  -  i  \sqrt{n_1} e^{i f_0(t)}    \right)     \psi_{-1}({\mathbf r})    \, .   
\label{bon2}
\end{multline}

In order to obtain $f_0(t)$ one can rewrite Eqs.~(\ref{ncmode2hn}) and
extract the evolution of $ f_0(t)$ from
$ \hbar\, \dot{\varphi_k} = ( n_{k-1} -n_{k}) N U_{\text{eff}} $. This
yields
\begin{equation}
f_0(t)= \frac{1}{\hbar} U_{\text{eff}} \Delta N\, t + f_0(0),
\label{f0}
\end{equation}
which in turns completely defines all the phase differences at any
time within the M model.  In particular, we have
$\varphi_0(t)= -f_0(t)$ and $\varphi_1(t)= f_0(t) -\pi $. We note that
for $n_0 = n_1$ the state (\ref{bon2}) coincides with the peculiar state
(\ref{gcuatrod}).

It is important to point out here, that the perfect agreement between
the results of GP equation and the M model observed in
Fig. \ref{fig:popu} is due to the proper definition of the onsite
interaction energy parameter
$U_{\mathrm{eff}}=2.27 \times10^{-3}\hbar\omega_r$
\cite{jezek13a,jezek13b}, instead of using the bare value
$U=3.16 \times10^{-3}\hbar\omega_r$. To illustrate such a difference,
we included in Fig. \ref{fig:popu} the evolution of $\varphi_1(t)$ for
$\Delta N =100$ using the bare parameter $U$.  Hence, we confirm the
accuracy on the calculation of the effective onsite interaction energy
parameter also in this dynamical regime.
    
To conclude we note that using Eq. (\ref{f0}) one can obtain the time
period for the phase differences,
\begin{equation}
T_{M} =  \frac{2 \pi  \hbar }{ U_{\text{eff}} \Delta N}.
\label{period}
\end{equation}
which turns to be also the time period of the persistent and
collective oscillation around the ring.

\subsubsection{ Angular momentum}

An additional evidence of this dynamical regime is reflected in the
time evolution of other observables. In particular, we shall show that
within the M model the angular momentum exhibits a sinusoidal behavior
as a function of time with a period $T_M$.

The expectation value per particle of a general observable
$\widehat{O}$, considering an arbitrary state in the M model is given
by
\begin{eqnarray}
\langle\widehat{O}\rangle &= &\sum_kn_k\langle w_k|\widehat{O}| w_k\rangle  \\
&+&
2 \sum_k\sqrt{n_kn_{k+1}}\Re\left[e^{i(\phi_k-\phi_{k+1})}\langle w_{k+1}|\widehat{O}| w_k\rangle\right]. \nonumber \\
\nonumber
\label{operadorgen1}
\end{eqnarray}
Since each well is equivalent to all others unless a discrete
rotation, we have
$\langle w_k|\widehat{O}|w_k\rangle=\langle w_0|\widehat{O}|w_0\rangle$
and
$\langle w_{k+1}|\widehat{O}|w_k\rangle=\langle w_1|\widehat{O}|w_0\rangle$
for all $k$. Hence, the expectation value becomes
\begin{eqnarray}
\langle\widehat{O}\rangle &= &  \langle w_0|\widehat{O}|w_0\rangle  \nonumber\\
&+&
2 \sum_k\sqrt{n_kn_{k+1}}\Re\left[e^{-i\varphi_{k+1}}\langle w_{1}|\widehat{O}|w_0\rangle\right]. \, 
\label{operadorgen2}
\end{eqnarray}
For the $z$-component of the angular momentum we have
$ \widehat{O}= \widehat{L}_z = -i \hbar
\frac{\partial}{\partial\theta} $ and then taking into account that
the localized states can be chosen as real functions, one obtains the
expectation value of angular momentum
\begin{equation}
\langle\widehat{L}_z\rangle = 
-2\hbar   \langle w_{1}| \frac{\partial}{\partial\theta}|w_0\rangle \sum_k\sqrt{n_kn_{k+1}} \sin{\varphi_{k+1}} \, .
\label{lz2}
\end{equation}
In a BON state with initial condition $f_0(0)=-\pi$, as
$ \sin{\varphi_{k}} =- \sin{f_0(t)}$ for every $k$, we can write
\begin{equation}
\langle\widehat{L}_z\rangle= 
 - 8\hbar\sqrt{n_0n_1}\langle w_1|\frac{\partial}{\partial\theta}|w_0\rangle \sin\left({\frac{1}{\hbar} U_{\text{eff}} \Delta N t}\right)
\label{4lz2}
\end{equation}
where the bracket involving the localized states is a negative number.
As expected, the period of this sinusoidal function is
$T_M$. Furthermore, one can see that the stationary state Eq.\
(\ref{orderwi0}), corresponding to $\Delta N=0$, yields a constant
angular momentum proportional to $\sin(f_0)$.

\subsubsection{ Underlying dynamics}

Although the population in each well remains completely fixed, the
order parameter evolves in time and exhibit spatial oscillations.  In
order to analyze such a dynamics we first investigate the evolution of
the density profile.  Using the BON state expression given by
Eq. (\ref{bon1}), the evolution of the density
$\rho_{M}({\mathbf r},t)= |\psi_{M} ({\mathbf r},t) |^2$ within the M
model is given by
\begin{multline}
\frac{\partial \rho_{M}({\mathbf r},t)}{\partial t} = 
2 \sqrt{n_0n_1}\dot{f_0}(t)\, \sin f_0(t)\,  \\
\times [{w}_0(\mathbf{r})-{w}_2(\mathbf{r})][{w}_1(\mathbf{r})-{w}_{-1}(\mathbf{r})], 
\label{gdro}
\end{multline}
with $ \hbar \dot{f_0}(t) = U_{\text{eff}} \Delta N $ (cf.\ Eq.\
(\ref{f0})).  Equation (\ref{gdro}) implies that
$\rho_{M}({\mathbf r},t)$ is approximately stationary within each
well, where the overlap between the WL functions of neighboring sites
is negligible.  Whereas the density variations are confined to the
inter-well regions or junctions where the localized states do overlap.
Moreover, one can infer the change of sign of
$\partial \rho_{M}({\mathbf r},t) / \partial t $ at the junctions by
analyzing Eq.\ (\ref{gdro}). One can thus conclude that particles
oscillate across both junctions of a given site without changing its
net population.  Furthermore, the maximum and minimum density
variations during the evolution occur at times $t_M$ and $t_m$ when
$f_0(t_M)= 0 $ and $f_0(t_m)= \pi$, respectively.

\begin{figure}
\begin{tabular}{cc}
\includegraphics[width=0.5\columnwidth,clip=true]{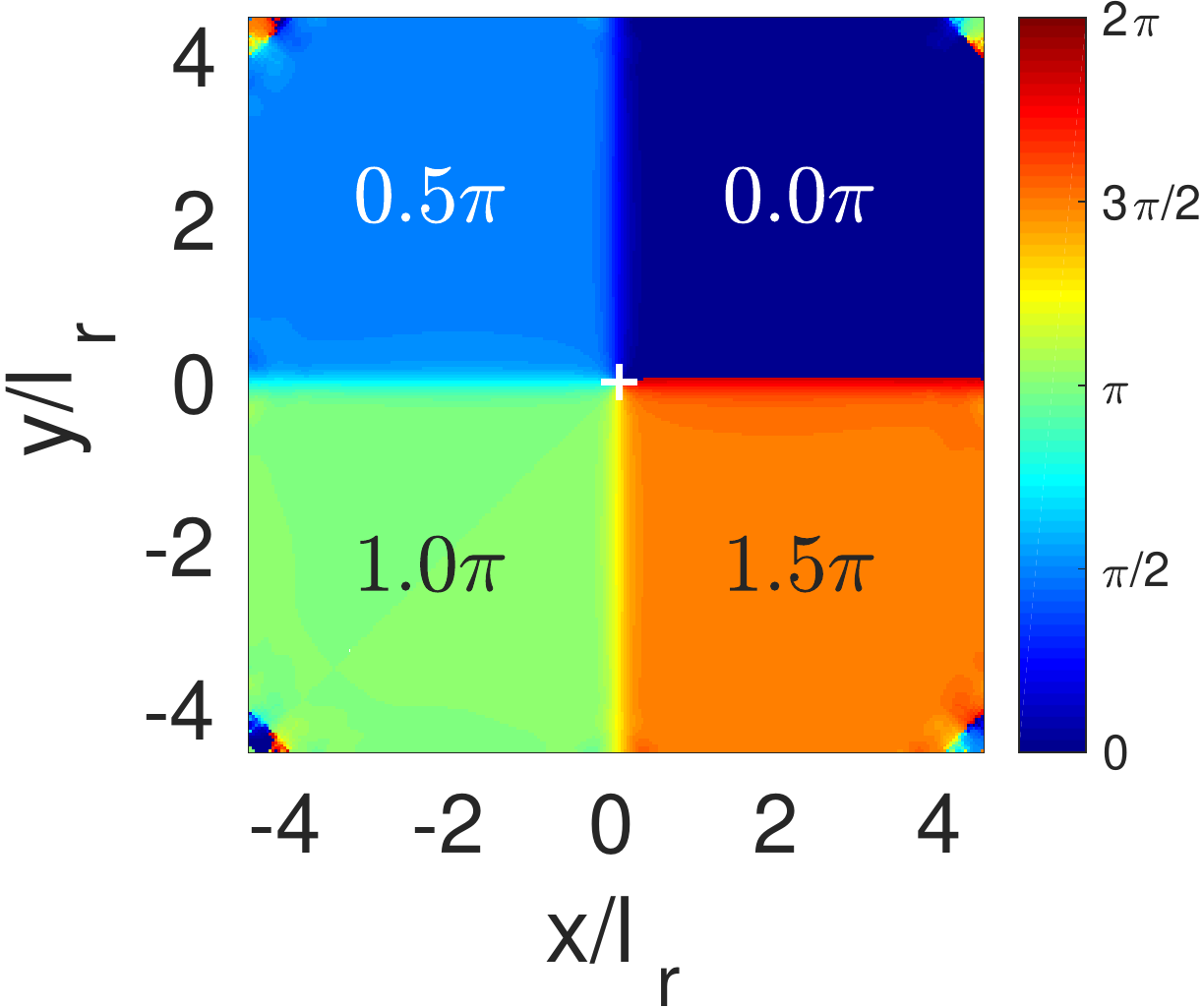}&
\includegraphics[width=0.5\columnwidth,clip=true]{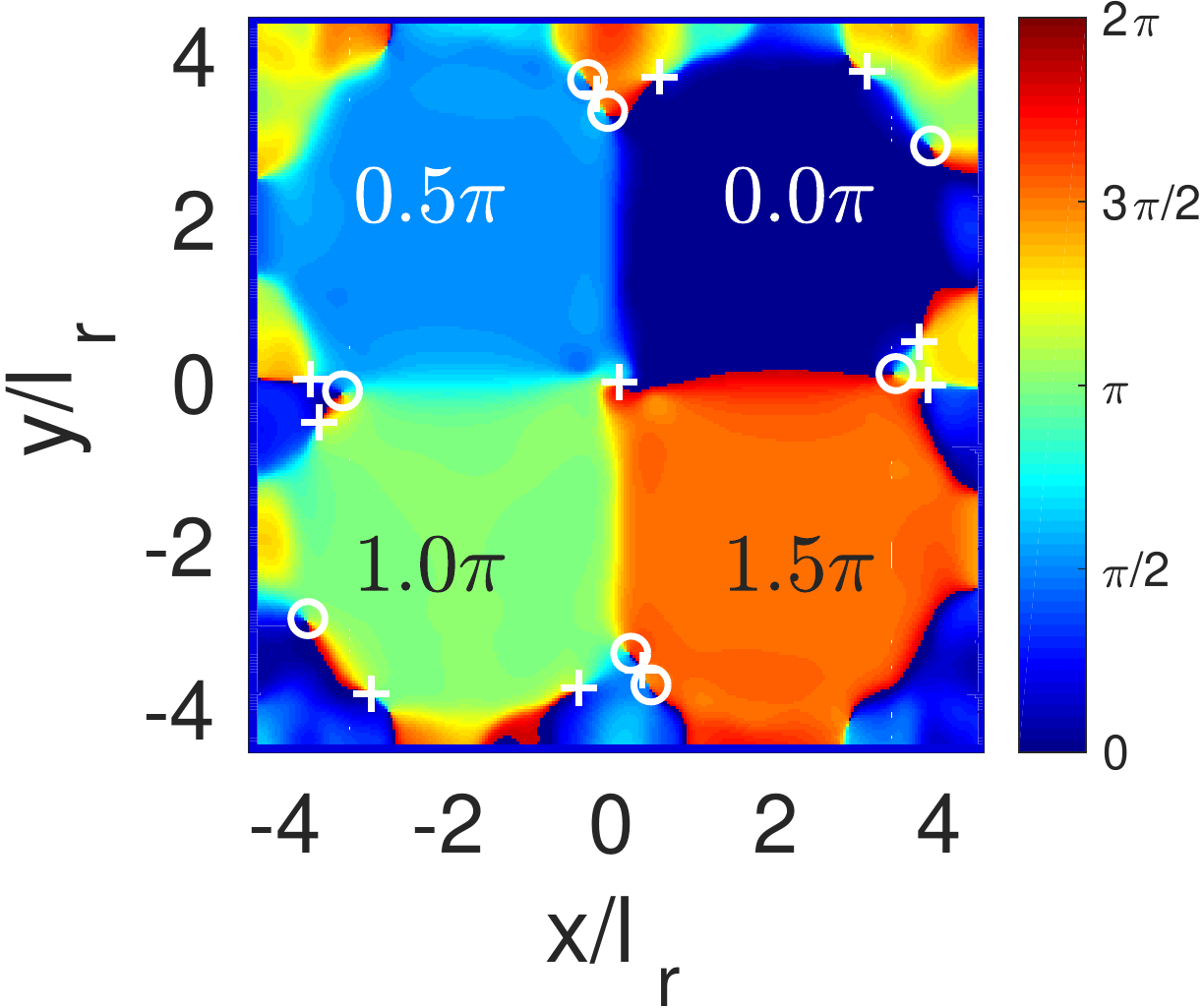} \\
\includegraphics[width=0.5\columnwidth,clip=true]{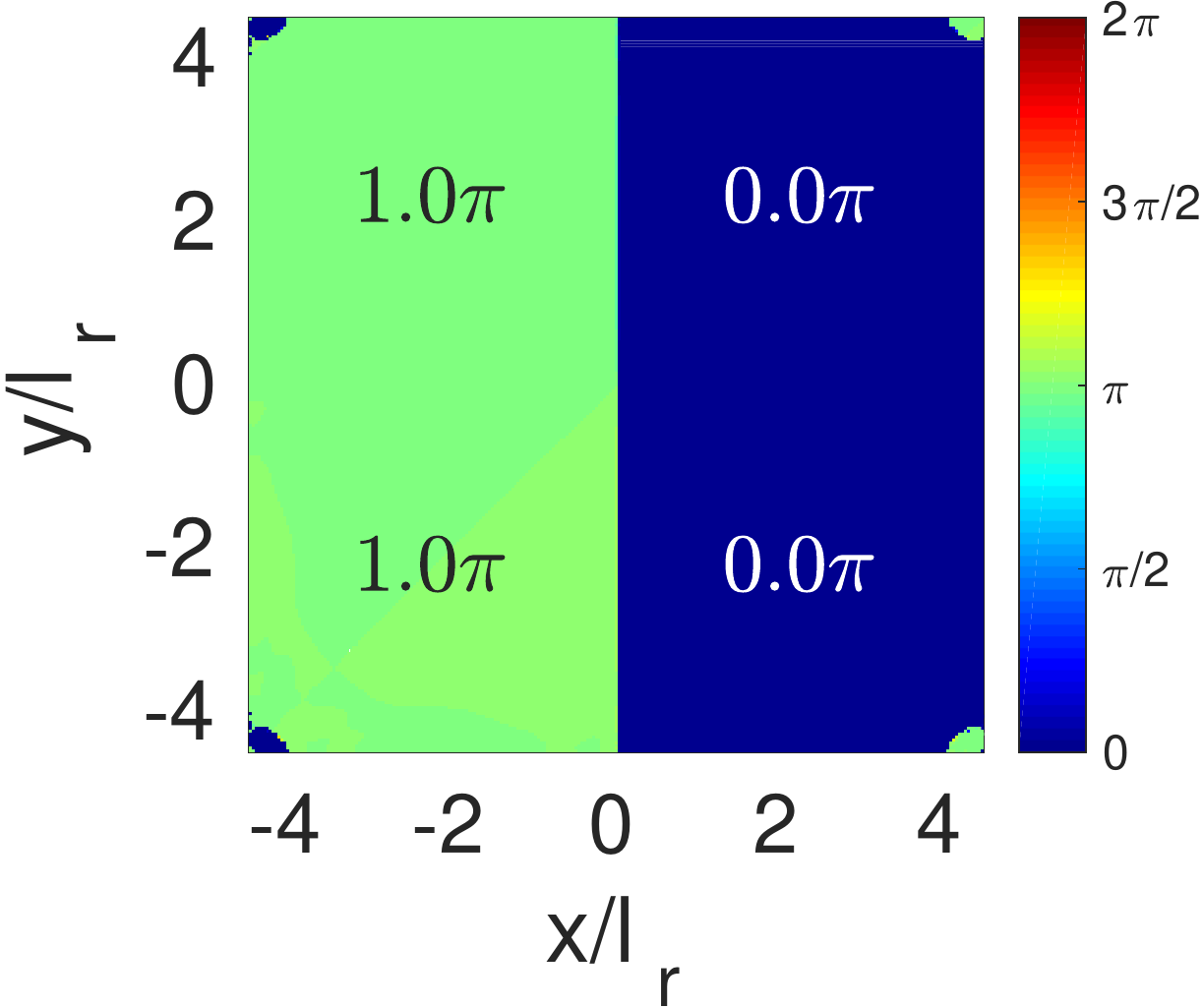}&
\includegraphics[width=0.5\columnwidth,clip=true]{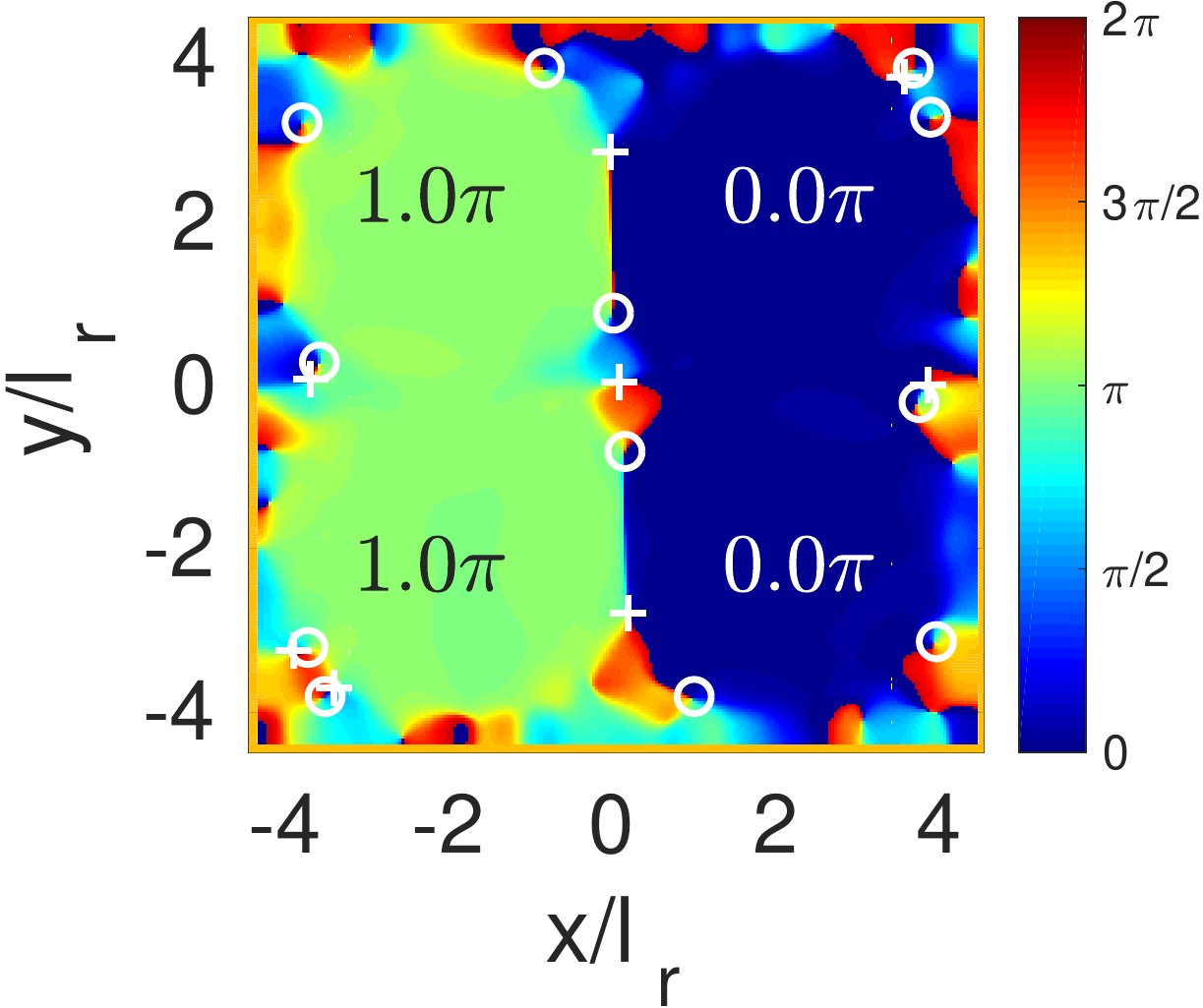} \\
\includegraphics[width=0.5\columnwidth,clip=true]{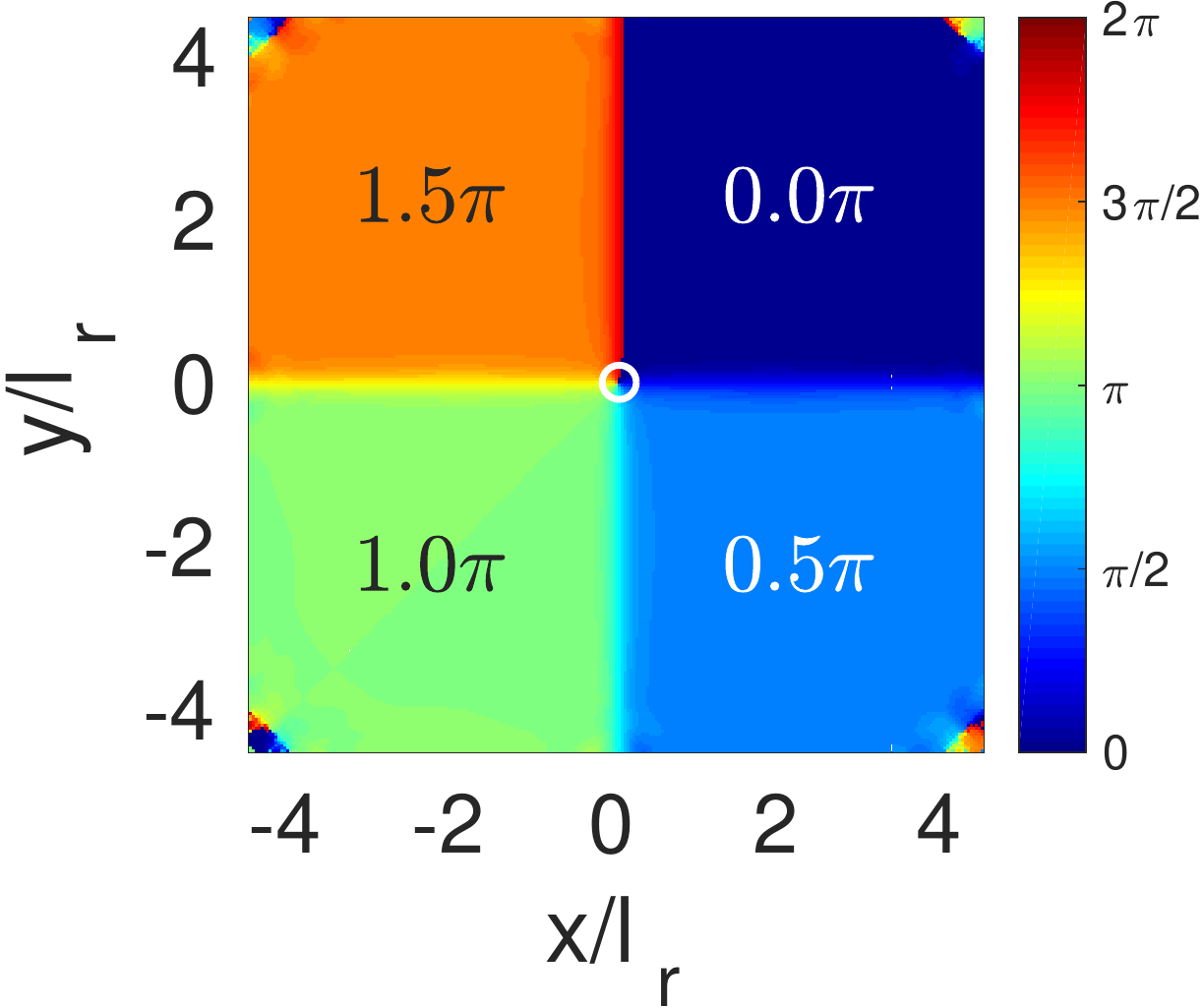}&
\includegraphics[width=0.5\columnwidth,clip=true]{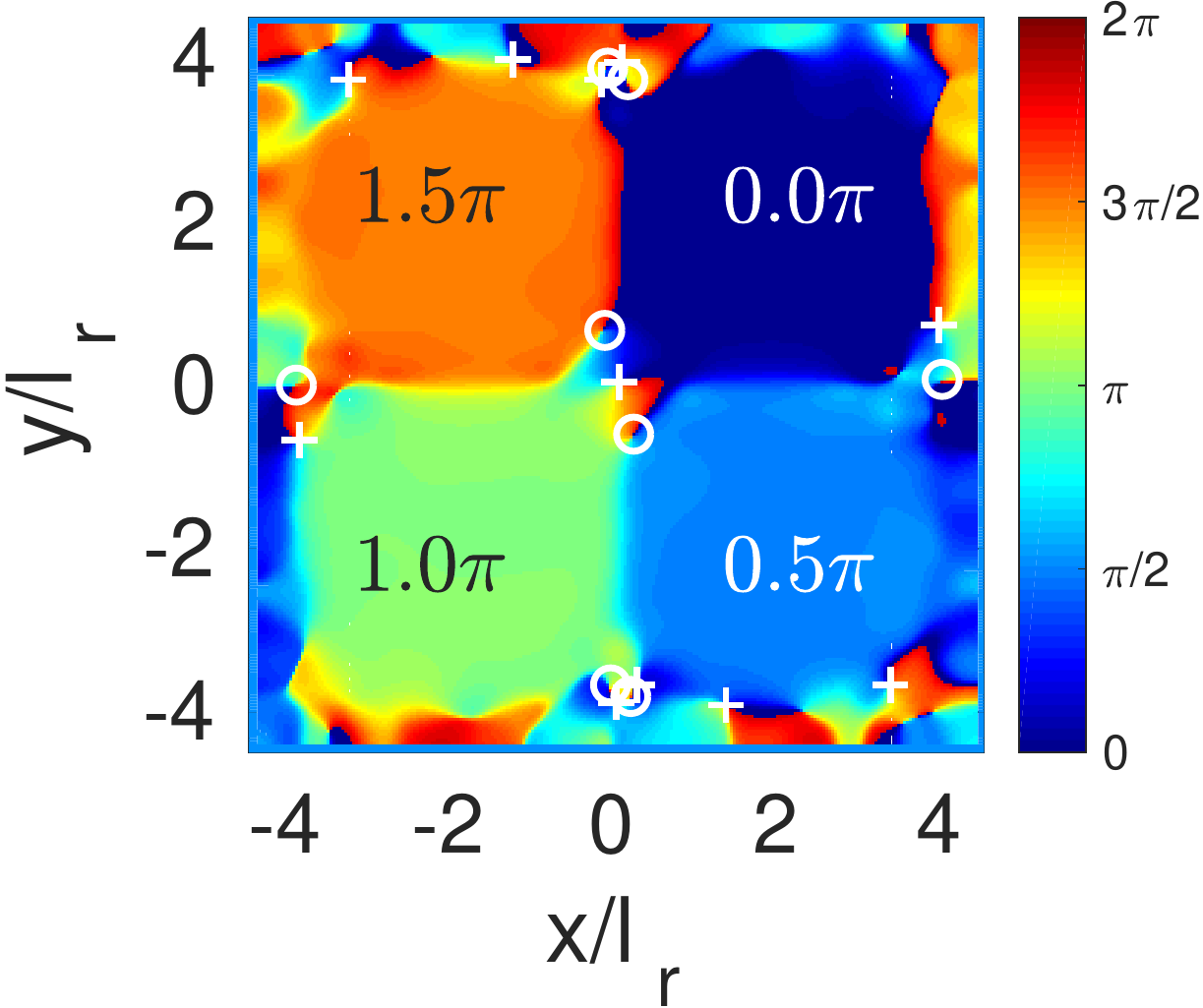} \\
\includegraphics[width=0.5\columnwidth,clip=true]{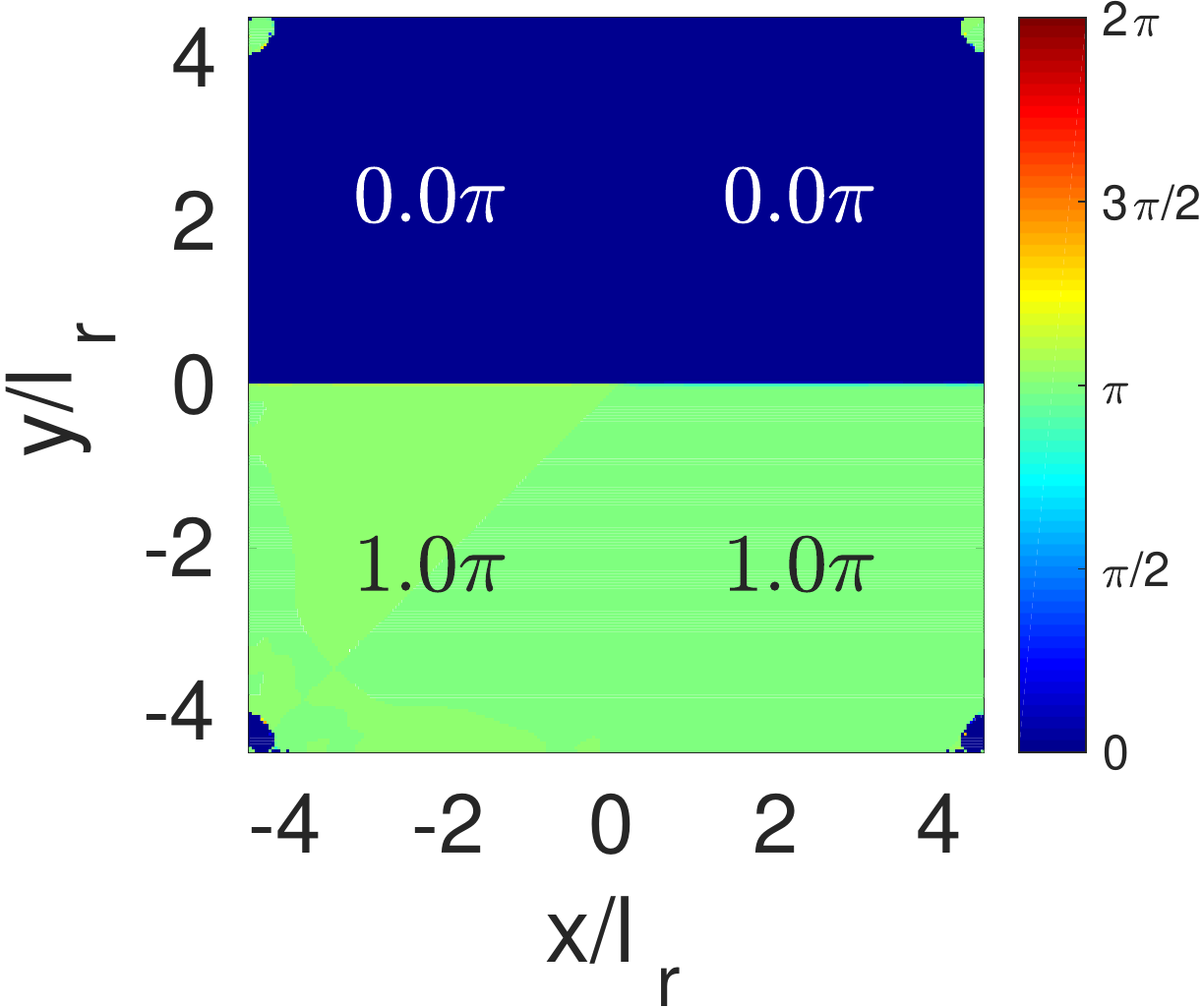}&
\includegraphics[width=0.5\columnwidth,clip=true]{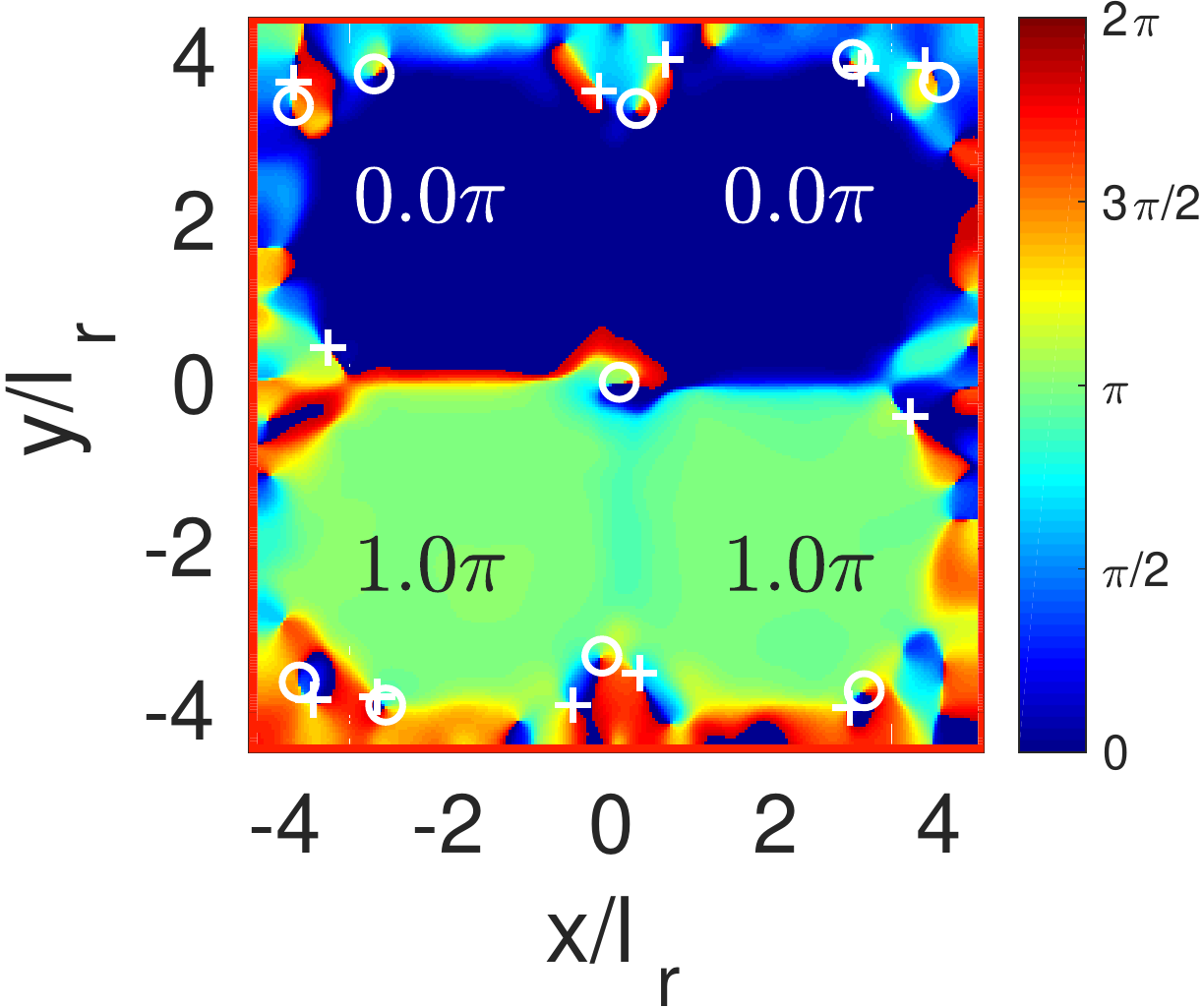} 
\end{tabular}
\caption{\label{fig:phase} Phase snapshots at the $z=0$ plane obtained
  from the M model (left column) and from GP simulations (right
  column) for $\Delta N= 200$. The coordinates $x$ and $y$ are divided
  by $ l_r=\sqrt{ \hbar/(m \omega_r)} $. The GP times, from top to
  bottom of the right (panel/column), correspond to :
  $ t_a=3.4 \omega_r^{-1} $, $ t_b=6.9 \omega_r^{-1}$,
  $ t_c=10.3 \omega_r^{-1} $, and $ t_d=13.7 \omega_r^{-1} $. The plus
  signs (empty circles) indicate the vortex (antivortex)
  locations. The indicated phase values correspond to the local phases
  evaluated at each site center.}
\end{figure}

The sense of the particles flow across the junctions can be read off
from the spatial profiles of the phases $\phi(\mathbf{r},t)$ of the
wavefunction. In Fig. \ref{fig:phase} we show snapshots of these
phases at several times in the $z=0$ plane obtained from both
$\psi_M ({\mathbf r},t)$ and $\psi_{\text{GP}}({\mathbf r},t) $ in the
left and right panels, respectively.  In both cases we have subtracted
a global phase $\phi_0(t)$ from $\phi(\mathbf{r},t)$ in order to
better observe the dynamics.  The initial condition is $ \Delta N=200$
and $f_0= - \pi $, which yield a multimode period
$ T_{M} \simeq 13.8 \omega_r^{-1}$, in sharp contrast to
$2 \pi \hbar / (U \Delta N)= 9.95 \omega_r^{-1} $ that would be
obtained with the bare onsite interaction.

In the left column, from top to bottom, we show the phase
$ \phi_M({\mathbf r},t) $ obtained from the order parameter
$\psi_M ({\mathbf r},t) $ for the aforementioned configurations at
several times: a) at $t=T_M/4$ ($f_0=-\pi/2$), there is a $\pi/2$
difference between neighboring sites and the velocity field
corresponds to that of a vortex with a phase gradient in the
counterclockwise direction, b) at $t=T_M/2$ ($f_0=0$), the phase
difference between the right and left sites is $\pi$, which
corresponds to a vanishing velocity field, c) at $t=3T_M/4$
($f_0=\pi/2$), there is a $-\pi/2$ difference between neighboring
sites, being the circulation clockwise as for an antivortex. Finally,
d) at $t=T_M$ ($f_0=\pi$), there is a $\pi$ phase difference between
the top and bottom sites.  In the figure we have marked with a plus
symbol and with an empty circle the presence of a vortex and an
antivortex, respectively.  For the M model, it can be seen that, in
the left column of Fig. \ref{fig:phase}, there exists a vortex and an
antivortex at the origin for the configurations: a) $t=T_{M}/4$ and c)
$t= 3/4 T_{M}$, in agreement with distributions described above. In
the model, the vortex (antivortex) remains fixed at $x=0,y=0$ during
the interval $ 0 < t < T_{M}/2$ ($ T_{M}/2 < t < T_{M}$).

On the other hand, on the right column of Fig. \ref{fig:phase} we show
phase snapshots obtained from full 3D GP simulations for times near
the four different situations previously discussed. We have observed
that the GP evolutions incorporate additional fluctuations and hence
the velocity circulation does not change exactly at quarters of the
period $T_M$. Moreover, the velocity field never vanishes, as the
change of its circulation is associated with a passage of vortices
instead of with the appearance of a nodal surface \cite{abad11}.
Nevertheless, as shown together in Fig. \ref{fig:phase}, the order
parameter from the M model is able to capture rather accurately the
spatial distribution of phases present in the exact GP dynamics.

In particular, from top to bottom in Fig.\ \ref{fig:phase}, we show
the results for the GP times: $ t_a=3.4 $$\omega_r^{-1} $, $ t_b=6.9
$$\omega_r^{-1}$, $ t_c=10.3 $$\omega_r^{-1} $, and $
t_d=13.7 $$\omega_r^{-1} $.  In each site we indicate the value of the
local phase $ \phi_{\text{GP}}({\mathbf r},t) $ evaluated at the
center of the corresponding well to be compared with that obtained in
the M model. It may be confirmed that at every time the phase
difference between alternated sites is always $\pi$ as predicted by
the model.

It becomes clear from the change of sign in the phase differences that
the velocity field is inverted near each half period, when the extreme
variations in the density at the junctions are achieved.  Except for
some fluctuations around such a transition, in the intermediate times
the total topological charge is conserved, whereas the number and the
position of the vortices may change.  In particular, in the third row
of the right column of Fig. \ref{fig:phase}, one vortex and two
antivortices are observed with a total negative charge of $-1$
instead of the single fixed antivortex predicted by the M model.

It is worthwhile to recall that the velocity circulation is quantized
along any closed curved inside the superfluid and, as established in
the celebrated Helmholtz-Kelvin theorem \cite{landau95}, it is
conserved during the evolution if the superfluid condition is not
broken \cite{bogdan03}. As a consequence, the value of the circulation
can only change when a vortex passes through the curve (phase slip) or
when the density goes to zero.

Although both the GP equation and the M model must obey the
Helmholtz-Kelvin theorem, the order parameter given by the multimode
model cannot predict the motion of vortices or the generation of
vortex-antivortex pairs, hence the change of the velocity field
circulation could be only provided through the appearance of nodal
surfaces. The nodal surfaces arise when the minimum in the local
density is achieved, i.e., at $f_0 =0, \pi$.  For example, at
$ f_0= 0$ the order parameter in Eq.\ (\ref{bon1}) reduces to
\begin{equation}
\psi_M ({\mathbf r}) =\sqrt{n_0}[ w_0 ({\mathbf r})- w_2({\mathbf r})] - \sqrt{n_1}
 [ w_1 ({\mathbf r})- w_{-1} ({\mathbf r})],
\label{orderwi0f0=0}
\end{equation}
which corresponds to the configuration b) on the left column of
Fig. \ref{fig:phase}. If all the populations were equal this condition
would lead to the $x=0$ plane.  In our case, the deviation from a
plane is due to the difference in the populations.  The intersection
of the nodal surface with the plane $z=0$ can be viewed in the graph
by the sharp $\pi$ change of the phase where the density goes to zero.
Similarly, one can obtain the nodal surfaces for $f_0=\pi$, which
corresponds to the configuration d).  In this case the curve where the
density goes to zero is around $ y=0$.

In contrast with the M model, the change of the velocity circulation
in the GP frame is produced by the dynamics of vortices passing
through the potential barriers and may include generation of
vortex-antivortex pairs.  In fact, we have observed that several
vortex-antivortex pairs may be spontaneously generated along the
barriers, thus simulating a density closer to that of the M model
nodal surface.  This active dynamics of vortices around the
transitions is produced in a timescale much smaller than $ T_M$ and
hence it is not possible to access the details of the vortex motion
within the present numerical precision.  As an illustration, we note
that the last time of the depicted GP snapshots is slightly smaller
than the $T_M$ period and there still exists an antivortex around the
center of the system.

\subsubsection{Velocity field circulation}

\begin{figure}
\begin{center}
\includegraphics[width=0.85\columnwidth,clip=true]{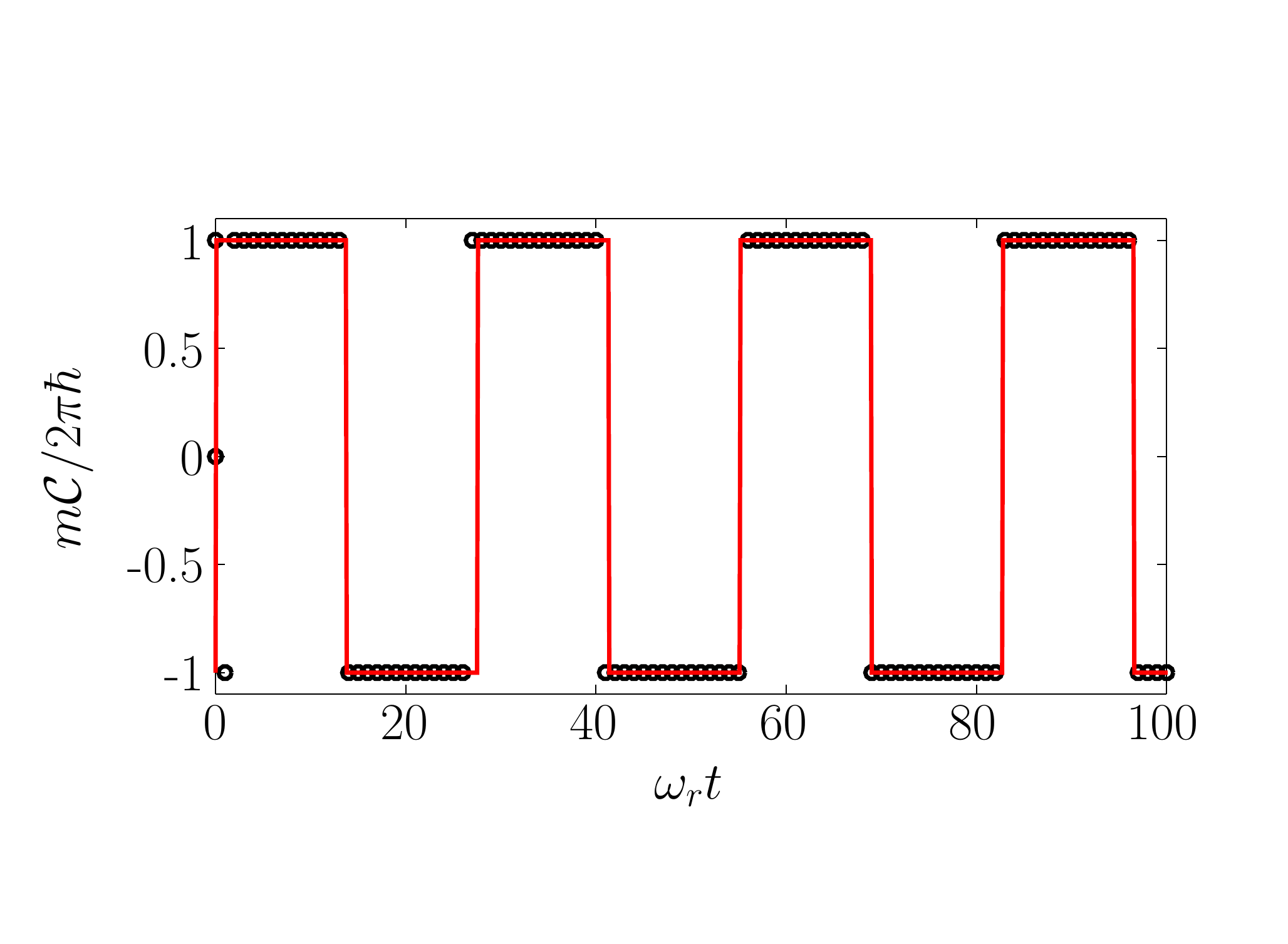}\\[-1.75cm]
\includegraphics[width=0.85\columnwidth,clip=true]{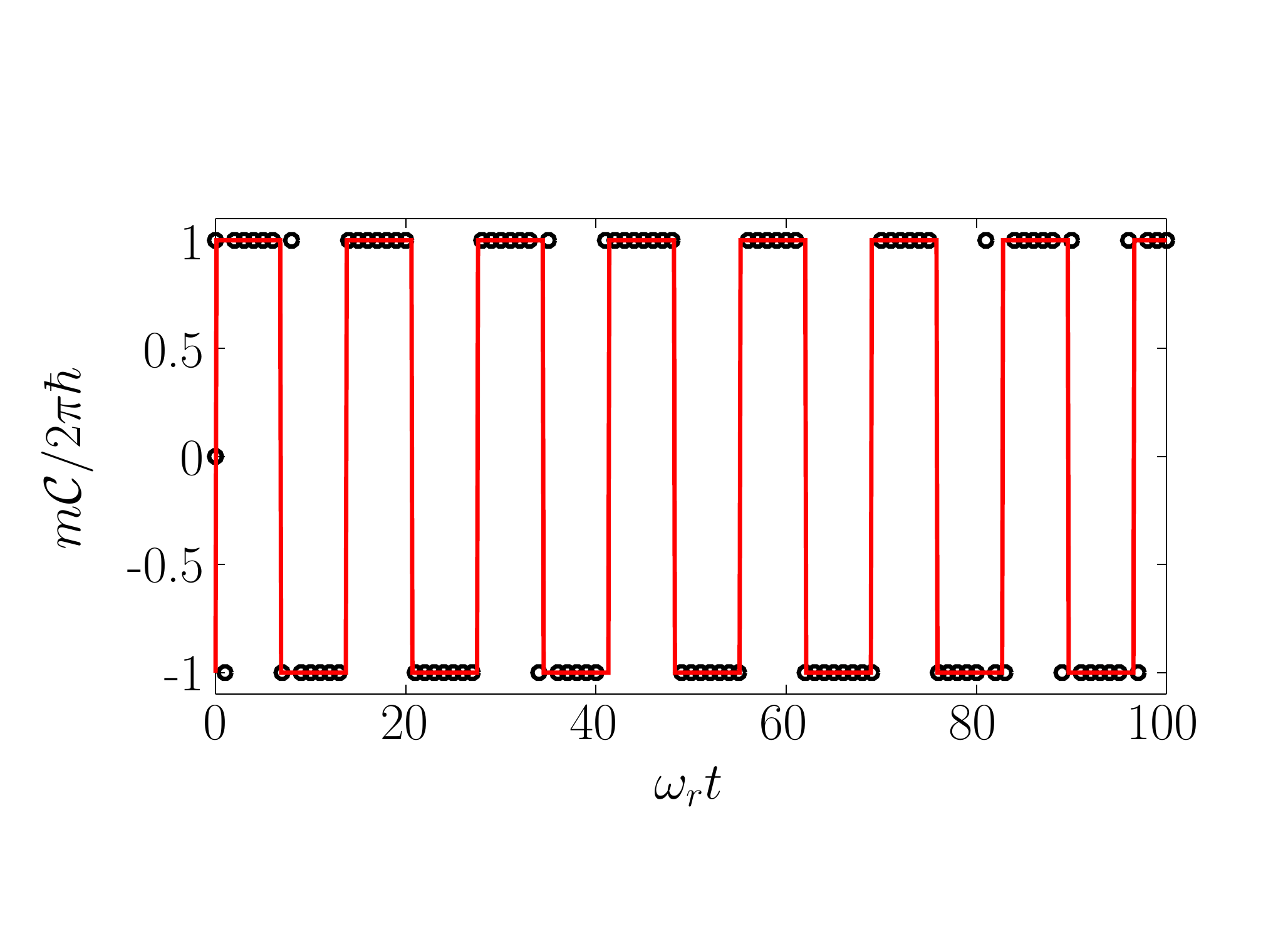}\\[-1.75cm]
\includegraphics[width=0.85\columnwidth,clip=true]{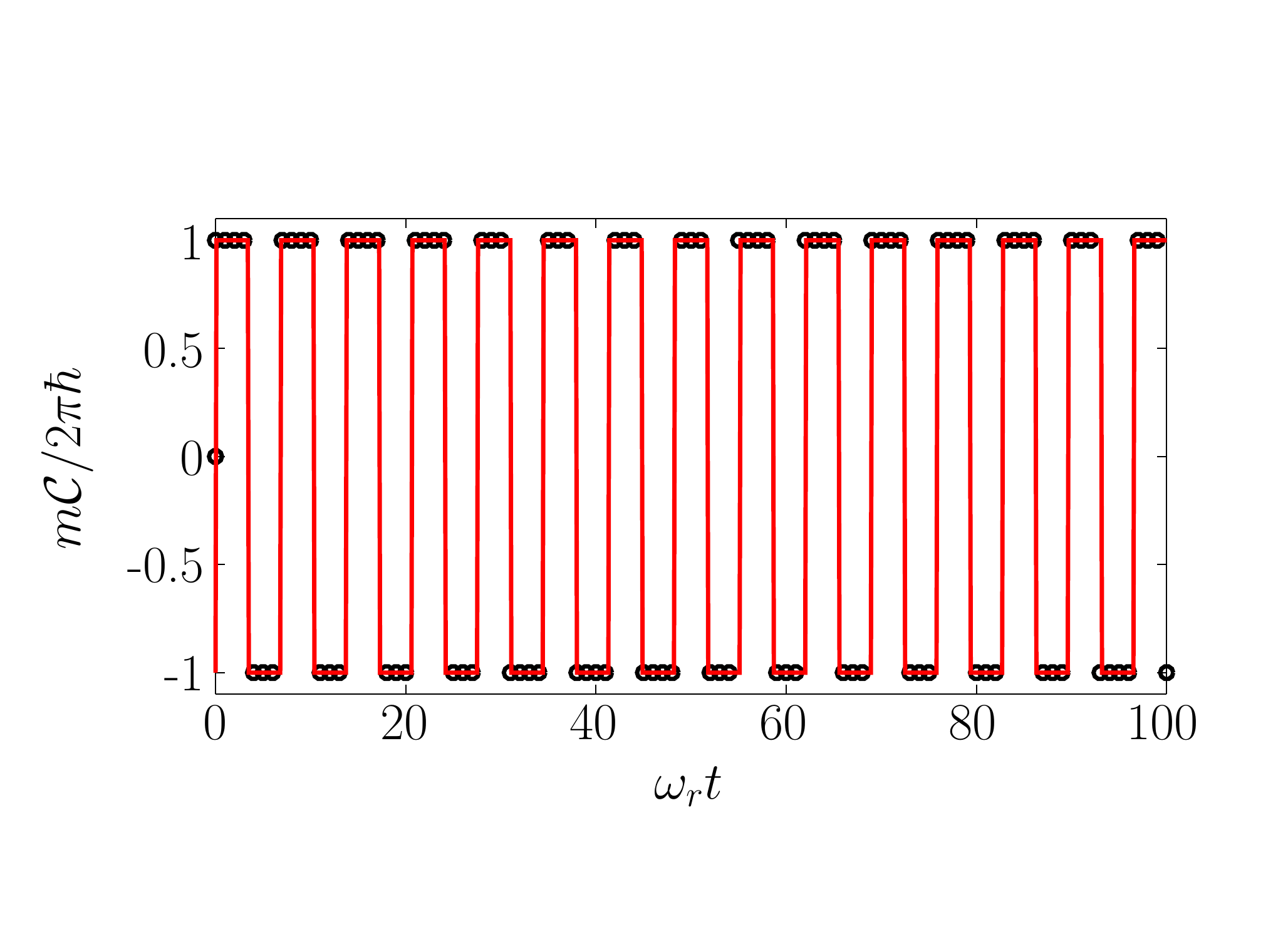}\vspace*{-1.25cm}
\end{center}
\caption{\label{fig:circu} (color online) Velocity field circulation
  $\mathcal{C}$ (in units of $2\pi\hbar/m$) as a function of time for $\Delta N=100, 200$, and
  $400$ (from top to bottom). The circles correspond to the GP results
  while the solid line to the results from the M model. The
  circulation was calculated along a square that connects the centers
  of the four sites in the $z=0$ plane.}
\end{figure}

Taking into account the previous findings for the multimode model one
can conclude that in one $T_M$ period the system passes through a
sequence of phases that yields an alternating velocity field
circulation between values $1$ and $-1$ along a curve that connects
the four wells. The transition between these two values occurs at
$ f_0= 0$ and $ f_0= \pi $ when the order parameter develops a nodal
surface.  In Fig. \ref{fig:circu} we show the velocity field
circulation $\mathcal{C}=\oint\mathbf{v}\cdot d\mathbf{r}$, as a
function of time using the M model and GP simulations. It may be seen
that the same behavior is observed with both approaches. The M model
is thus able to reproduce the behavior of the circulation although the
details of the internal vortex dynamics is lost. In the GP dynamics
the change of circulation is caused by the motion of vortices together
with the creation or annihilation of vortex-antivortex pairs.
Signatures of such a vortex dynamics could be observed in
Fig. \ref{fig:phase} where we have shown the phases around the
transition.  Another evidence of a vortex dynamics can also be
visualized in the middle panel of Fig. \ref{fig:circu}, where an
additional change of sign is produced near the transition.

\subsubsection{Stability analysis}
\begin{figure}
\includegraphics[width=0.9\columnwidth,clip=true]{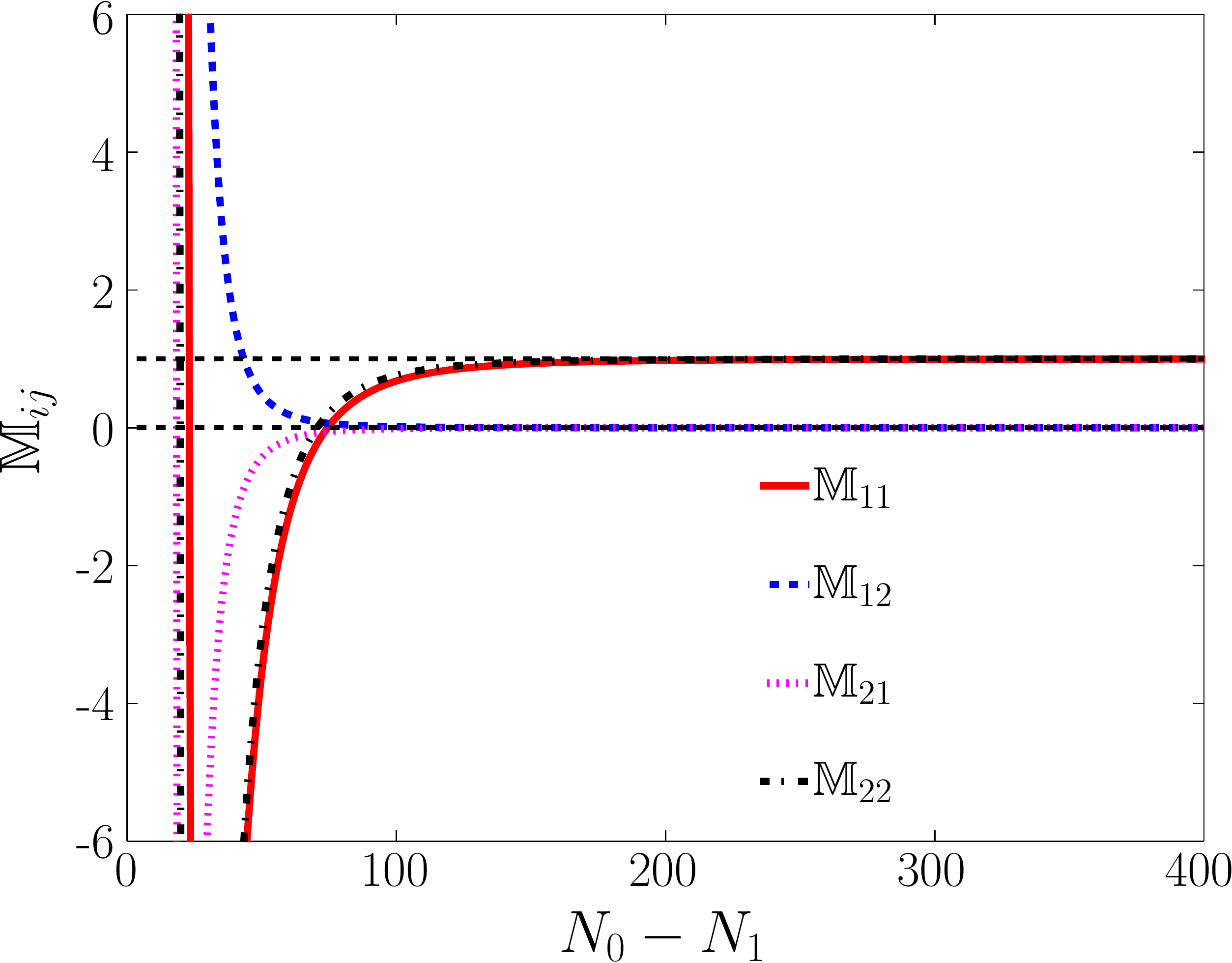}
\caption{\label{fig:Mono} Selected elements of the Monodromy matrix
  $\mathbb{M}_{ij}$ as functions of the particle imbalance $N_1-N_2$
  of BON states with $f_0(0)=\pi$. The horizontal dashed lines mark
  the values $\mathbb{M}_{ij}=0, 1$.}
\end{figure}

In this section we investigate the stability of the BON states by
means of a Floquet analysis \cite{Floquet,mauro4p} of the multimode
dynamical equations. This analysis is based on the characterization of
the linear dynamics around its periodic orbits. In our case, the BON
states are periodic solutions with constant populations $n_i(t)=n_i$
and linear phase differences
$\varphi_i(t)= \varphi_i^0 + (-1)^{i+1} 2\pi\, {t}/{T_M}$ where
$\varphi_i^0$ fulfill the relation:
$\varphi_0^0=-f_0(0), \varphi_1^0=f_0(0)-\pi, \varphi_2^0=2\pi-f_0(0),
\varphi_{-1}^0=f_0(0)-\pi$. The linearization of the dynamics around
these states yield the non-autonomous system
\begin{equation}
\frac{d\bm{\delta}}{dt}=\mathbb{A}\left[n_i(t),\varphi_i(t)\right]\Bigr|_{\text{BON}}\cdot\bm{\delta}(t)
\label{eq:dEq}
\end{equation}
where $\bm{\delta}$ is a vector comprising both density and
phase-difference fluctuations. As the BON states correspond to symmetric
initial populations with peculiar phases, it is natural to consider as
variables $\bm{\delta}$ the departures from a symmetric case, namely,
we define
\begin{align}
  \delta_1 &= (n_0-n_2)/2, &\delta_2= (n_1-n_{-1})/2, \\
  \delta_3 &= (\varphi_0-\varphi_2)/2 + \pi, &\delta_4=(\varphi_1-\varphi_{-1})/2.
\end{align}
Given that the matrix $\mathbb{A}$ has a period $T_M$, the linearized
dynamics can be characterized by the so-called Monodromy matrix
$\mathbb{M}$ which contains the change of $\bm{\delta}$ after one
period, i.e., $\mathbb{M}\cdot\bm{\delta}(0)=\bm{\delta}(T_M)$. The
matrix is built from the solutions of Eq.~(\ref{eq:dEq}) with
canonical initial conditions evaluated at
$T_M$ \cite{Floquet,mauro4p}. In Fig.\ \ref{fig:Mono} we depict
elements of $\mathbb{M}$ showing the effect of an initial population
fluctuation. The orbits are regular if the perturbed system remains
near the initial one after a period. This happens when
$\mathbb{M}_{ii}\simeq 1$ and $\mathbb{M}_{i\neq j}\simeq 0$. On the
contrary, when the fluctuations are enhanced
($|\mathbb{M}_{ij}|\gg 1$), the orbits are unstable. This may be observed
in Fig.\ \ref{fig:Mono} for low imbalances.

For the peculiar stationary states ($\Delta N=0$), the linear system is
time-independent and the problem reduces to a straightforward
diagonalization of $\mathbb{A}$ to obtain the excitation frequencies
$\tilde{\omega}$ corresponding to the Bogoliubov collectives modes in
the case of the full GPE. The four frequencies are found to be
\begin{equation}
  \tilde{\omega} = \pm \sqrt{F^2\,\cos^2f_0 \pm K \frac{N U_{\text{eff}}}{2}\cos f_0 -(K+F)^2} 
\label{eq:omegaEx}
\end{equation}
where $K=2J+F$. As $N U_{\text{eff}}\gg K, F$, the most stable
frequency for a given system is attained for peculiar states with
$f_0=\pm\pi/2$ which in turn yield an imaginary frequency
$\tilde{\omega}^2 = -(K+F)^2$. Therefore all $f_0$ give rise to
dynamically unstable peculiar states. The stability of stationary
vortex states ($f_0=\pm \pi/2$) has been previously investigated in
\cite{Paraoanu2003} for circular arrays of BECs, finding that only
states with circulation below $N_c/4$ are stable.

\subsubsection{Proposed experimental test}  
The correct preparation of BON states requires a special sequence of
phases and symmetric initial populations ($n_{k}=n_{k+1}$). While the
common approach to experimentally measure both of them is by means of
TOF and absorption images, the simple dynamics of BON states offers an
alternative way to confirm its correct realization using TOF images
only. Given that the relative phases among neighboring sites are
revealed in the interference patterns during the TOF expansion
\cite{albiez05}, it could be verified that they obey the peculiar
sequence of phases at all times.  According to Eq.~(\ref{f0}), in this
case $f_0(t)$ must be a linear function whose slope $\gamma$ relates
to the particle imbalance as
\begin{equation}
\Delta N=\frac{\hbar\gamma}{U_{\mathrm{eff}}},
\label{imbalance}
\end{equation}
which might probe to be a more accurate measure than the direct
estimate from absorption images. On the other hand, since
Eq. (\ref{imbalance}) requires the use of $U_{\mathrm{eff}}$ instead
of the bare $U$, it may also serve to confirm its numerical value.  By
using $U$ the relative error on the imbalances could be as large as of
order 20-30\%, depending on the number of particles \cite{mauro17}.

Due to the experimental uncertainty, absorption images may not be able
to reveal a slightly broken symmetry of the population
configuration. However, the evolution of the phases will depart from
linearity and will not be determined by the single function $f_0$.

\section{ Extension to  larger number of wells}\label{larger}

It is possible to extend the peculiar and BON states to larger number
of wells provided the sequence of phases
$ ...0,f_0 -\pi, \pi, f_0, 0...$ is repeated $l=N_c/4$ times around
the ring lattice, and the populations alternate between two values,
with $n_{2k}=n_0 $ and $n_{2k+1}=n_1 $.  This is only possible when
the number of wells are multiples of 4. Taking into account these
conditions in Eq. (\ref{orderparameter}) and using Eq. (\ref{wannier})
to eliminate the WL functions $w_k$, we can write the following
BON order parameter in terms of GP stationary states,
\begin{eqnarray}
  \psi_M ({\mathbf r})   &=&   \frac{\sqrt{N_c}}{2} \left[  (   \sqrt{n_0}   +  i  \sqrt{n_1}  e^{i  f_0(t)}   )  
  \psi_{\frac{N_c}{4}} ({\mathbf r})\right. \nonumber \\
  &+& \left. ( \sqrt{n_0}  -  i  \sqrt{n_1} e^{i f_0(t)} )  \psi_{-\frac{N_c}{4}}({\mathbf r}) \right]    \, .   
\label{gbon2}
\end{eqnarray}
It is straightforward to show that $f_0(t)$ still obeys
Eq.~(\ref{f0}), and thus the corresponding time period is also given
by Eq. (\ref{period}). Therefore, the analysis performed in the
previous section can be repeated using the same procedure, including
the Floquet theory.  However, for these configurations the velocity
field circulation alternates between $ \pm N_c/4$ and the number of
nodal surfaces at each half period is equal to $l$.  Equation
(\ref{gbon2}) shows that an arbitrary linear combination of
$\psi _{\pm \frac{N_c}{4}}$ leads to the BON dynamics. For example,
even though for $N_c=8$ it is not possible to generate the BON
dynamics with a linear combination of the degenerate $\psi_{\pm 1}$
states; any linear combination of $\psi _{\pm 2}$ will indeed give
rise to a BON dynamics.

Using Eq. (\ref{lz2})  the mean value of the $z$-component of the angular momentum is given by,
\begin{equation}
\langle L_z \rangle (t) =   2  N_c  \hbar  \langle  w_1  | \frac{\partial}{\partial \theta} |  w_{0} \rangle   
  \sqrt{n_0 n_{1}}     \sin ( f_0(t) ).
\label{lznc}
\end{equation} 

If we let $n_0=n_1$ then we obtain the most general quasi-stationary states described in section IV.A:
\begin{eqnarray}
  \psi_M ({\mathbf r})   &=&   \frac{1}{2} \left[  (   1   +  i    e^{i  f_0}   )  
  \psi_{\frac{N_c}{4}} ({\mathbf r})\right. \nonumber \\
  &+& \left. ( 1  -  i   e^{i f_0} )  \psi_{-\frac{N_c}{4}}({\mathbf r}) \right]   \, .   
\label{peculiar general}
\end{eqnarray}
that satisfies 
\begin{multline}
 \left[ -\frac{ \hbar^2}{2m} \nabla^2 + V_{\text{trap}} +
g \, |\psi_M (\mathbf{r})|^2   \right] \psi_M ({\mathbf r})  =   \mu_{\frac{N_c}{4}}  \,  \psi_M ({\mathbf r}) \\
 -  g N \cos(f_0) \Im\left( \psi_{\frac{N_c}{4}}^2({\mathbf r})\right) \left[ \Re ( \psi_{\frac{N_c}{4}} ) - \Im (\psi_{\frac{N_c}{4}}) e^{i f_0} \right]  
\label{gened2}
\end{multline}
Since 
\begin{eqnarray}
\Im \left( \psi_{\frac{N_c}{4}}^2\right) &=& \frac{1}{N_c}\sum _{k,k'}w_k w_{k'} \sin \left [\frac{\pi}{2}(k+k')\right ] \nonumber \\
\label{cuasiN}
\end{eqnarray}
the states Eq. (\ref{peculiar general}) can be regarded as
quasi-stationary solutions of the GP equation when the $w_k$ are
well localized functions.

\section{ Summary and concluding remarks}\label{sum} 

We have studied a particular dynamical regime of a Bose-Einstein
condensate in a ring-shaped lattice which possesses a set of states
with fixed number of particles in each site and a simple dynamics in
their phases. The same distribution of phases along the sites that
gives rise to such nonstationary states has been shown to generate a
continuous family of stationary points in the phase space of the
multimode model. Such peculiar states have constant nonzero angular
momentum, when all the populations are equal, and include two states
that correspond to exact GP stationary solutions.

We have shown that the nonlinearity of the GP equation governs the
dynamics within this regime and that it is responsible for the
population blocking in the nonstationary states.  In contrast to the
self-trapping phenomenon this effect does not possess a lower bound
for the population imbalance.

We have studied the time evolution of BON states using both the
multimode model and the three-dimensional GP equation finding an
excellent agreement in the populations in each site and in their phase
differences.  This accuracy was possible due to the inclusion of the
effective interaction energy parameter instead of the bare one. Even
though the multimode model was unable to account for the motion of
individual vortices and the creation/annihilation of vortex-antivortex
pairs, it was demonstrated that it correctly predicts the evolution of
the velocity circulation and angular momentum, characterizing this
regime as a persistent current oscillating around the lattice.

By performing a Floquet stability analysis of the blocked populations
states, we have verified that their dynamics is regular for the
particle imbalances here considered. In a four-well system these
states could, in principle, be experimentally achieved by initially
manipulating the position of the potential barriers in order to have
different populations or by using an elliptic trap in the $ (x,y)$
plane with their axis forming a $\pi/4$ angle during a short time and
then reverting the potential to a circular harmonic trap.  A simple
way to produce the initial distribution of phases would be to start
with the same state as we have used in our numerical
calculations. This could be achieved by illuminating half of the
condensate (e.g., $x>0$ ) with an additional laser for a period of
time until it develops a $\pi$ phase difference between the half
spaces $x>0$ and $x<0$. However, any other initial distribution of
phases seems feasible using a Spatial Light Modulator (SLM)
\cite{slm1,slm2,slm3}, and hence also the whole family of stationary
M-model states could be directly generated.  Furthermore, given that
all the phases loose their dependence on a single linear function as
soon as the symmetric condition on the site populations is lifted,
these states could be first tested to adjust the population in
alternate wells with arbitrary imbalances. A second phase imprinting
application could then be used to generate the desired state.  Since
the BON states present a simple analytical form for the phase
difference between neighboring sites, it could also allow to measure
the initial population imbalance by means on interference patterns in
TOF images, rather than absorption images.

\appendix*

\section{  Parameters} \label{parameters} 

The multimode model parameters are defined by

\begin{equation}
J= -\int d^3{\bf r}\,\, w_0 ({\bf r}) \left[
-\frac{ \hbar^2 }{2 m}{\bf \nabla}^2  +
V_{\text{trap}}({\bf r})\right]  w_1({\bf r}),
\label{jota0}
\end{equation}
\begin{equation}
U= g \int d^3{\bf r}\,\,  w_0^4({\bf r}), \quad \text{and}
\label{U0}
\end{equation}
\begin{equation}
F= -  N \,g\int d^3{\bf r}\,\,  w_0^3({\bf r})
 w_1 ({\bf r}).
\label{jotap0}
\end{equation}

Together with the calculation of these parameters by the preceding
definitions we have followed the alternative method outlined in
Ref. \cite{jezek13b} which involves directly the energies of the GP
stationary states.  Both approaches have proven to yield values equal
in less than one percent. We note that we have disregarded the
parameter that involves products of neighboring densities because, for
the present system, its contribution turned out to be negligible.

This work was supported by CONICET and Universidad de
Buenos Aires through grants PIP 11220150100442CO and UBACyT
20020150100157BA, respectively.

\end{document}